\documentclass[twocolumn]{aastex631}
\usepackage{float}
\usepackage{amsmath}

\graphicspath{{./}{figures/}}
\shorttitle{Compact PSBs at $z \sim 1.2$}
\shortauthors{Zhang et al.}

\begin{document}

\title{DESI Massive Post-Starburst Galaxies at $\mathbf{z\sim1.2}$ have compact structures and dense cores}

\author[0000-0001-6454-1699]{Yunchong Zhang} 
\affiliation{Department of Physics and Astronomy and PITT PACC, University of Pittsburgh, Pittsburgh, PA 15260, USA}

\author[0000-0003-4075-7393]{David J. Setton}\thanks{Brinson Prize Fellow}
\affiliation{Department of Astrophysical Sciences, Princeton University, Princeton, NJ 08544, USA}

\author[0000-0002-0108-4176]{Sedona H. Price}
\affiliation{Department of Physics and Astronomy and PITT PACC, University of Pittsburgh, Pittsburgh, PA 15260, USA}

\author[0000-0001-5063-8254]{Rachel Bezanson}
\affiliation{Department of Physics and Astronomy and PITT PACC, University of Pittsburgh, Pittsburgh, PA 15260, USA}

\author[0000-0002-3475-7648]{Gourav Khullar}
\affiliation{Department of Physics and Astronomy and PITT PACC, University of Pittsburgh, Pittsburgh, PA 15260, USA}

\author[0000-0001-8684-2222]{Jeffrey A. Newman}
\affiliation{Department of Physics and Astronomy and PITT PACC, University of Pittsburgh, Pittsburgh, PA 15260, USA}

\author{Jessica Nicole Aguilar}
\affiliation{Lawrence Berkeley National Laboratory, 1 Cyclotron Road, Berkeley, CA 94720, USA}

\author[0000-0001-6098-7247]{Steven Ahlen}
\affiliation{Physics Dept., Boston University, 590 Commonwealth Avenue, Boston, MA 02215, USA}

\author[0000-0001-8085-5890]{Brett H. Andrews}
\affiliation{Department of Physics and Astronomy and PITT PACC, University of Pittsburgh, Pittsburgh, PA 15260, USA}

\author{David Brooks}
\affiliation{Department of Physics \& Astronomy, University College London, Gower Street, London, WC1E 6BT, UK}

\author{Todd Claybaugh}
\affiliation{Lawrence Berkeley National Laboratory, 1 Cyclotron Road, Berkeley, CA 94720, USA}

\author[0000-0002-1769-1640]{Axel de la Macorra}
\affiliation{Instituto de F\'{\i}sica, Universidad Nacional Aut\'{o}noma de M\'{e}xico,  Cd. de M\'{e}xico  C.P. 04510,  M\'{e}xico}

\author[0000-0002-5665-7912]{Biprateep Dey}
\affiliation{Department of Physics and Astronomy and PITT PACC, University of Pittsburgh, Pittsburgh, PA 15260, USA}

\author{Peter Doel}
\affiliation{Department of Physics \& Astronomy, University College London, Gower Street, London, WC1E 6BT, UK}

\author{Enrique Gaztañaga}
\affiliation{Institut d'Estudis Espacials de Catalunya (IEEC), 08034 Barcelona, Spain}
\affiliation{Institute of Cosmology and Gravitation, University of Portsmouth, Dennis Sciama Building, Portsmouth, PO1 3FX, UK}
\affiliation{Institute of Space Sciences, ICE-CSIC, Campus UAB, Carrer de Can Magrans s/n, 08913 Bellaterra, Barcelona, Spain}

\author[0000-0003-3142-233X]{Satya Gontcho A Gontcho}
\affiliation{Lawrence Berkeley National Laboratory, 1 Cyclotron Road, Berkeley, CA 94720, USA}

\author[0000-0002-5612-3427]{Jenny E. Greene}
\affiliation{Department of Astrophysical Sciences, Princeton University, Princeton, NJ 08544, USA}

\author{Stephanie Juneau}
\affiliation{NSF NOIRLab, 950 N. Cherry Ave., Tucson, AZ 85719, USA}

\author{Robert Kehoe}
\affiliation{Department of Physics, Southern Methodist University, 3215 Daniel Avenue, Dallas, TX 75275, USA}

\author[0000-0003-3510-7134]{Theodore Kisner}
\affiliation{Lawrence Berkeley National Laboratory, 1 Cyclotron Road, Berkeley, CA 94720, USA}

\author[0000-0002-7613-9872]{Mariska Kriek}
\affiliation{Leiden Observatory, Leiden University, P.O. Box 9513, NL-2300 AA Leiden, The Netherlands}

\author[0000-0001-6755-1315]{Joel Leja}
\affiliation{Department of Astronomy \& Astrophysics, The Pennsylvania State University, University Park, PA 16802, USA}
\affiliation{Institute for Computational \& Data Sciences, The Pennsylvania State University, University Park, PA 16802, USA}
\affiliation{Institute for Gravitation and the Cosmos, The Pennsylvania State University, University Park, PA 16802, USA}

\author[0000-0003-4962-8934]{Marc Manera}
\affiliation{Departament de F\'{i}sica, Serra H\'{u}nter, Universitat Aut\`{o}noma de Barcelona, 08193 Bellaterra (Barcelona), Spain}
\affiliation{Institut de F\'{i}sica d’Altes Energies (IFAE), The Barcelona Institute of Science and Technology, Campus UAB, 08193 Bellaterra Barcelona, Spain}

\author[0000-0002-1125-7384]{Aaron Meisner}
\affiliation{NSF NOIRLab, 950 N. Cherry Ave., Tucson, AZ 85719, USA}

\author{Ramon Miquel}
\affiliation{Instituci\'{o} Catalana de Recerca i Estudis Avan\c{c}ats, Passeig de Llu\'{\i}s Companys, 23, 08010 Barcelona, Spain}
\affiliation{Institut de F\'{i}sica d’Altes Energies (IFAE), The Barcelona Institute of Science and Technology, Campus UAB, 08193 Bellaterra Barcelona, Spain}

\author[0000-0002-2733-4559]{John Moustakas}
\affiliation{Department of Physics and Astronomy, Siena College, 515 Loudon Road, Loudonville, NY 12211, USA}

\author[0000-0001-7145-8674]{Francisco Prada}
\affiliation{Instituto de Astrof\'{i}sica de Andaluc\'{i}a (CSIC), Glorieta de la Astronom\'{i}a, s/n, E-18008 Granada, Spain}

\author{Graziano Rossi}
\affiliation{Department of Physics and Astronomy, Sejong University, Seoul, 143-747, Korea}

\author[0000-0002-9646-8198]{Eusebio Sanchez}
\affiliation{CIEMAT, Avenida Complutense 40, E-28040 Madrid, Spain}

\author{Michael Schubnell}
\affiliation{Department of Physics, University of Michigan, Ann Arbor, MI 48109, USA}
\affiliation{University of Michigan, Ann Arbor, MI 48109, USA}

\author[0000-0002-2949-2155]{Małgorzata Siudek}
\affiliation{Institute of Space Sciences, ICE-CSIC, Campus UAB, Carrer de Can Magrans s/n, 08913 Bellaterra, Barcelona, Spain}

\author[0000-0003-3256-5615]{Justin Spilker}
\affiliation{Department of Physics and Astronomy and George P. and Cynthia Woods Mitchell Institute for Fundamental Physics and Astronomy, Texas A\&M University,
4242 TAMU, College Station, TX 77843-4242, USA}

\author{David Sprayberry}
\affiliation{NSF NOIRLab, 950 N. Cherry Ave., Tucson, AZ 85719, USA}

\author[0000-0002-1714-1905]{Katherine A. Suess}\thanks{Hubble Fellow}
\affiliation{Kavli Institute for Particle Astrophysics and Cosmology and Department of Physics, Stanford University, Stanford, CA 94305, USA}

\author[0000-0003-1704-0781]{Gregory Tarl\'{e}}
\affiliation{University of Michigan, Ann Arbor, MI 48109, USA}

\author[0000-0002-6684-3997]{Hu Zou}
\affiliation{National Astronomical Observatories, Chinese Academy of Sciences, A20 Datun Rd., Chaoyang District, Beijing, 100012, P.R. China}

\author{DESI Collaboration}
\begin{abstract}

Post-starburst galaxies (PSBs) are young quiescent galaxies that have recently experienced a rapid decrease in star formation, allowing us to probe the fast-quenching period of galaxy evolution. In this work, we obtained HST WFC3/F110W imaging to measure the sizes of 171 massive ($\mathrm{log(M_{*}/M_{\odot})\sim\,11)}$ spectroscopically identified PSBs at $1<z<1.3$ selected from the DESI Survey Validation Luminous Red Galaxy sample. This statistical sample constitutes an order of magnitude increase from the $\sim20$ PSBs with space-based imaging and deep spectroscopy. We perform structural fitting of the target galaxies with \texttt{pysersic} and compare them to quiescent and star-forming galaxies in the 3D-HST survey. We find that these PSBs are more compact than the general population of quiescent galaxies, lying systematically $\mathrm{\sim\,0.1\,dex}$ below the established size-mass relation. However, their central surface mass densities are similar to those of their quiescent counterparts ($\mathrm{\,log(\Sigma_{1\,kpc}/(M_{\odot}/kpc^2))\sim\,10.1}$). These findings are easily reconciled by later ex-situ growth via minor mergers or a slight progenitor bias. These PSBs are round in projection ($b/a_{median}\sim0.8$), suggesting that they are primarily spheroids, not disks, in 3D. We find no correlation between time since quenching and light-weighted PSB sizes or central densities. This disfavors apparent structural growth due to the fading of centralized starbursts in this galaxy population. Instead, we posit that the fast quenching of massive galaxies at this epoch occurs preferentially in galaxies with pre-existing compact structures.

\end{abstract}


\section{Introduction} \label{sec:intro}
It is well established that galaxies can be divided into two categories in terms of their star-formation activities: star-forming galaxies, whose star-formation rates (SFRs) scale with their stellar masses; and quiescent galaxies, which form very few stars relative to their existing stellar mass \citep[e.g.,][]{Blanton&Moustakas2009, Wuyts.etal.2011,Whitaker.etal.2012b}. Furthermore, as revealed by extra-galactic surveys in recent years \citep[e.g.,][]{vanderWel.etal.2014,Straatman.etal.2015,Mowla.etal.2019,Kawinwanichakij.etal.2021,Martorano.etal.2024}, quiescent galaxies are more compact than their star-forming counterparts at fixed stellar masses across all epochs up to $z\sim 4$. Such a bimodal distribution in star-formation activity and structure among galaxy populations implies a connection between the structural transformation and the shutting off of star formation in those galaxies. 

It has become increasingly evident that the timescales over which galaxies quench can be subdivided into two modes, which dominate the transformations at different cosmic epochs. Galaxies can shut down slowly by gradually assembling their stellar masses and exhausting their gas reserve, which is common at low redshifts or they can decrease in star formation rates within just millions of years. The latter ``fast mode" typically occurs after an extreme starburst and is more prevalent at higher redshifts \citep[e.g.,][]{Wu.etal.2018, Rowlands.etal.2018, Belli.etal.2019, Park.etal.2024}. Simulations have shown that gas-rich dissipative mergers or interactions can induce both central starburst and compaction of structure in galaxies \citep[e.g.,][]{Wellons.etal.2015,Zolotov.etal.2015,Zheng.etal.2020}, which can be later followed by rapid quenching if paired with active galactic nucleus (AGN) feedback or starburst-driven winds \citep{Hopkins.etal.2005}. Alternatively, a number of compact star-forming galaxies exist at high redshifts \citep{Barro.etal.2013,Barro.etal.2014a,Barro.etal.2014b,Williams.etal.2014,Barro.etal.2017} and their number densities decrease over cosmic time, along with the increase in the quiescent galaxy number density \citep{vanDokkum.etal.2015}. This suggests that morphological transformation could have already occurred during the star-forming phase for some galaxies. These plausible scenarios point out ways to link morphological compaction to the cessation of star formation among the fast-quenching galaxies. However, it is still under debate what exact physical mechanisms affect the rapid suppression of star formation and whether those physical mechanisms are universally dominant. Hence, placing observational constraints on the properties of galaxies on the fast quenching track is vital to our understanding of how exactly their quenching process relates to their morphological transformation. 

Post-starburst galaxies (PSBs), which are also known as ``K+A" galaxies, provide us with a unique view of this fast-quenching process. Typically, these galaxies experienced a dramatic drop in star formation in the most recent $1 \, \rm Gyr$ and are not currently forming stars. As a result, their spectra are dominated by the flux from late-type B and type A stars \citep{Dressler.etal.1983,Zabludoff.etal.1996}. In common practice, PSBs are selected by methods that aim to identify spectral energy distribution (SED) features that represent such unique stellar population composition, which includes searching for the existence of strong Balmer absorption along with a lack of nebular emission lines \citep[e.g.,][]{Goto.etal.2005,French.etal.2015,Wu.etal.2018,Chen.etal.2019}, ``K+A" template fitting \citep[e.g.,][]{Pattarakijwanich.etal.2016}, photometric selection methods (UVJ color space \citep[e.g.,][]{Whitaker.etal.2012a,Belli.etal.2019,Suess.etal.2020}; super colors \citep[e.g.,][]{Wild.etal.2014,Wild.etal.2016,Wilkinson.etal.2021}, or unsupervised machine learning techniques \citep[e.g.,][]{Meusinger.etal.2017}). A detailed review of PSBs selection methods can be found in \cite{French.etal.2021}.

To date, structural studies of PSBs have yielded interesting insights into their evolutionary histories. At intermediate redshifts ($z\sim0.7$), studies have shown that PSBs have systematically smaller half-light radii than the general population of quiescent galaxies at the same mass and the same epoch \citep[e.g.,][]{Yano.etal.2016,Almaini.etal.2017,Maltby.etal.2018,Wu.etal.2018,Suess.etal.2020,Setton.etal.2022} while having similar central densities as quiescent galaxies \citep{Setton.etal.2022}. In addition, spatially-resolved slits or integral field unit spectra have revealed younger stellar population in the center than outskirts in PSBs at both intermediate redshifts \citep{D'Eugenio.etal.2020b} ($z\sim0.7$) and local universe \citep{Chen.etal.2019,Wu.etal.2021}. In consonance with these findings in PSBs, starburst components of small sizes ($\sim 0.1\,\rm kpc$) have been observed in massive starburst galaxies \citep{Sell.etal.2014, Diamond-Stanic.etal.2021}. These results jointly favor central starbursts as the mechanism of the morphological transformation in the fast quenching path. On one hand, the studies of the most massive PSBs ($\rm log(M_{*}/M_{\odot}) > 10$) at intermediate redshifts found no significant color or age gradients \citep{Maltby.etal.2018,Setton.etal.2020,Suess.etal.2020}. On the other hand, negative color gradients are found in both typical star-forming and quiescent galaxies \citep{Mosleh.etal.2017,Suess.etal.2019a,Suess.etal.2019b,Mosleh.etal.2020,Suess.etal.2021}, suggesting that the half-mass radii are smaller than the half-light radii in these galaxies. If PSBs evolve from those typical star-forming galaxies, the central starburst can drastically lower the mass-to-light ratio in the center of the galaxy and potentially explain the flattening of color gradients. By incorporating quenching time inferred from spectroscopy, \cite{Setton.etal.2022} examine the PSB size as a function of $t_{q}$. They found that PSBs at $z \sim 0.7 $ were compact and their sizes hardly evolved relative to the coeval quiescent galaxies in the first $1\, \rm Gyr$ since quenching. This result suggests that the starburst occurred at a spatial scale comparable to the half-light radii of the PSB progenitors.  Taken together, the physical conditions necessary to produce fast quenching may be different for the progenitors of PSBs in different mass regimes at earlier epochs. For those massive PSB galaxies, their star-forming progenitors would already have been compact. 

Since the PSB number densities only start to increase at $z>1$ \citep{Whitaker.etal.2012b,Wild.etal.2016,Clausen.etal.2024} and fast-quenching likely dominates the earlier universe, we need to further study PSBs at earlier cosmic times to better understand the role of fast-quenching in galaxy evolution. Firstly, high signal-to-noise rest-frame optical spectra are needed to pinpoint the time since quenching ($t_q$) while high-resolution imaging is required for precise characterization of morphology. Given the rarity of massive ($\rm log(M_{*}/M_{\odot}) > 10.8$) PSBs at $z>1$ \citep{Wild.etal.2016}, only a handful of objects have been studied with deep spectroscopy at these redshifts in the full CANDELS/3D-HST fields \citep[e.g.,][]{Belli.etal.2017,Belli.etal.2019,Kriek.etal.2019}. However, the recent Dark Energy Spectroscopic Instrument \citep{DESI.2016,DESI.2022} Survey Validation dataset, which was released publicly as a part of the Early Data Release \citep{DESI_EDR,DESI.2024} and contains 1-5 hour integrations of $\sim$20000 luminous red galaxies (LRGs) from $z=0.4-1.3$, has enabled us to select PSBs by their Balmer absorption strength ($\rm H_{\delta, a} > 4 \AA$) and resulted in an order of magnitude increase in the number of spectroscopically confirmed post-starburst galaxies at $z>1$ \citep{Setton.etal.2023}. While the DESI Legacy Survey imaging \citep{Dey.etal.2019} that this sample was selected from has insufficient resolution to perform structural analysis, HST data (HST SNAP 17710, PI: D. Setton) for 171 out of the ($\rm H_{\delta, a} > 4 \AA$) 409 galaxies at $z>1$ are recently obtained and allow us to perform joint analysis of the structures and star formation histories for a previously unprecedented sample.

In this work, we study the sizes, structures, and merger signatures of 171 PSBs selected from $1<z<1.3$ DESI Luminous Red Galaxies, using WFC3/F110W imaging from HST in conjunction with star-formation history derived from DESI spectroscopy. This represents an order of magnitude increase in the sample size of massive ($\rm log(M_{*}/M_{\odot}) > 10.8$) recently quenched galaxies at $z>1$. The statistical power of such a sample size enables us to robustly characterize these galaxies' morphology and test whether there's any structural transformation within the population of fast-quenching galaxies in the first $1\,\rm Gyr$ since they became quiescent. In Section \ref{sec:data}, we describe the target selection and reference sample from the 3D-HST survey. We describe our methodology for size measurements and structural analysis in Section \ref{sec:method} and the results of this work in Section \ref{sec: analysis}. In Section \ref{sec: discussion}, we discuss the results and their implication to the fast quenching pathway of galaxy evolution. Throughout this paper, we assume 
 a flat $\rm \Lambda CDM$ cosmology with $ \Omega_{\Lambda} = 0.69$, $ \Omega_{m} = 0.31$ and $H_{0} = \rm 67.66 \, km\,s^{-1}\,Mpc^{-1}$ as reported in \cite{Planck18} and quote AB magnitudes. All reported effective radii ($R_{e}$) are measurements of the semi-major axis and are not circularized. 

\section{Data} \label{sec:data}
\subsection{The DESI Survey Validation Luminous Red Galaxy Spectroscopic Sample} \label{sec:PSB Selection}

\begin{figure*}[!ht]
    \centering
    \includegraphics[width = \textwidth]{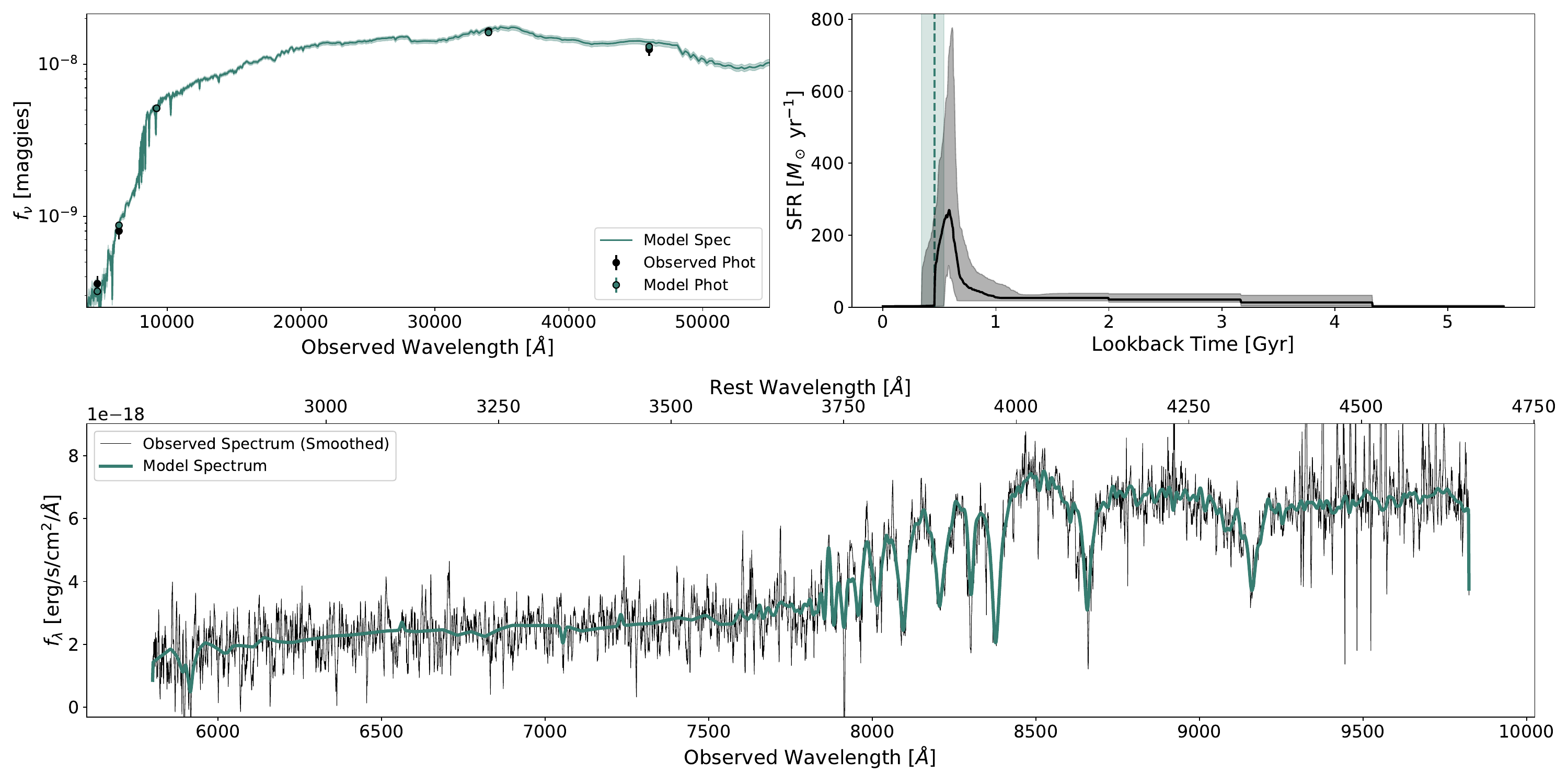}
    \caption{An example of spectro-photometric fitting for one of the PSB galaxies. In the upper left panel, we show the best-fitting models (green) and the observed photometry ({\it g/r/z/W1/W2}, black). In the upper right panel, we show the median (black solid line) and 68\% confidence interval (gray region) SFH. The time since quenching is defined as the look-back time where sSFR drops below the threshold of $\rm log(sSFR [yr^{-1}]) < -10$. We over-plot the median (green dashed line) and 68\% confidence interval (light green region) of the quenching time. In the lower panel, we show the smoothed observed spectrum (5-pixel boxcar smoothed, black) and the best-fitting model spectrum (green). }
    \label{fig:sSFR}
\end{figure*}

The galaxies in this work were observed as a part of the Luminous Red Galaxy component of the DESI Survey Validation Sample \citep{Zhou.etal.2020,Zhou.etal.2023}, which became public as a part of the DESI Early Data Release \citep{DESI_EDR,DESI.2024}. We note that the sample selection of DESI Validation Sample LRG is limited by a z-band magnitude cut of $\rm z_{fiber}<21.6$ and biased towards brighter objects \citep{Zhou.etal.2023} at $\rm z>0.8$. The Dark Energy Spectroscopic Instrument (DESI) is a robotic, fiber-fed, highly multiplexed spectroscopic surveyor that operates on the Mayall 4-meter telescope at Kitt Peak National Observatory \citep{DESI.2022}. DESI, which can obtain simultaneous spectra of almost 5000 objects over a $\sim 3^{\circ}$ field \citep{DESI.etal.2016b,Silber.etal.2023,Miller.etal.2023}, is currently conducting a five-year survey of about a third of the sky. This campaign will obtain spectra for approximately 40 million galaxies and quasars \citep{DESI.2016}.

The goal of the Dark Energy Spectroscopic Instrument (DESI) is to determine the nature of dark energy through the most precise measurement of the expansion history of the universe ever obtained \citep{Levi.etal.2013}. DESI was designed to meet the definition of a Stage IV dark energy survey with only a 5-year observing campaign. Forecasts for DESI \citep{DESI.2016} predict a factor of approximately five to ten improvement on the size of the error ellipse of the dark energy equation of state parameters $w_{0}$ and $w_{a}$ relative to previous Stage-III experiments. 

The sheer scale of the DESI experiment necessitates multiple supporting software pipelines and products, which include significant imaging from the DESI Legacy Imaging Surveys (\citealp{zou.etal.2017,Dey.etal.2019}; Schlegel et al. 2024 in preparation), an extensive spectroscopic reduction pipeline \citep{Guy.etal.2023}, a template-fitting pipeline to derive classifications and redshifts for each targeted source (Redrock; Bailey et al. 2024 in preparation, and for the special case of QSOs, \citealp{Brodzeller.etal.2023}), a pipeline to assign fibers to targets (Raichoor et al. 2024 in preparation), a pipeline to tile the survey and to plan and optimize observations as the campaign progresses \citep{Schlafly.etal.2023}, an online exposure time calculator (Kirkby et al. 2024 in preparation), a pipeline to select targets for spectroscopic follow-up (desitarget; \citealp{Myers.etal.2023}), and Maskbit to generate ``clean" photometry \citep{Moustakas.etal.2023}. 

We initially selected 409 galaxies from the DESI Validation Sample by their $\rm H_{\delta}$ spectral indices ($\rm H_{\delta, a} > 4 \AA$) at $1<z<1.3$ as our primary sample. In order to strictly select the galaxies that are quiescent, we apply an additional selection criteria that requires specific star formation rate (sSFR; SFR divided by stellar mass) to be lower than $10^{-10} yr^{-1}$ for our primary sample. This sSFR threshold was chosen such that new stellar mass formed is insignificant relative to the existing stellar mass, which is commonly used to define passive galaxies in literatures \citep[e.g.,][]{Wellons.etal.2015,Salim.etal.2018}. To enable the secondary selection and better quantity the star-formation histories (SFH) of these galaxies, we make use of the joint spectrophotometric spectral energy distribution (SED) fits from \citet{Setton.etal.2023}, which modeled the spectro-photometric data for the entire Survey Validation LRG sample at $0.4<z<1.3$. The data included the Milky Way extinction-corrected g/r/z/W1/W2 photometry and the spectra covering $\rm 5800 \AA < \lambda_{obs} < 9824 \AA$, in which the resolution $R \, (\lambda/\Delta\lambda)$ varies from $\sim 3200$ to $5100$. Briefly, this spectro-photometric fitting used the Bayesian stellar population synthesis code \texttt{Prospector} \citep[V1.2,][]{Leja.etal.2017,Johnson.etal.2021}, assuming a non-parametric SFH optimized for fitting post-starburst galaxies \citep{Suess.etal.2022}. The fits assumed a \cite{Chabrier.etal.2003} initial mass function and used the mass–metallicity prior described in \cite{Leja.etal.2019}.  We refer the reader to \cite{Setton.etal.2023} for further details.

 In Figure \ref{fig:sSFR}, we show an example of the spectro-photometric fitting. Using the derived SFH, we calculated the fraction of mass formed in the most recent $1\, Gyr$ ($f_{1Gyr}$) and the time since quenching ($t_q$). We define $t_q$ as the time elapsed since the last time the sSFR dropped below $10^{-10} yr^{-1}$.  We report the uncertainties of $t_q$ for each object as the time when 16 or 84 percent of the posterior drop below the aforementioned sSFR threshold. We note that 32 of the total 171 $\rm H_{\delta, a} > 4 \AA$ galaxies are determined to be currently star-forming based on their sSFR. These galaxies are excluded from the subsequent analysis but are included in the catalog that accompanies this paper.

\subsection{HST WFC3/F110W Imaging}\label{sec:HST}

Because the resolution (pixel scale $\sim 0.''262$) and seeing of the ground-based DESI Legacy Survey Imaging \citep[][Schlegel et al. 2024 in preparation]{zou.etal.2017, Dey.etal.2019} is insufficient to measure robust sizes in compact quiescent galaxies at $z>1$ \citep[which we expect to have sizes $\leq1$", see e.g.,][]{Almaini.etal.2017, Maltby.etal.2018}, we instead utilize HST/WFC3 F110W imaging that was measured as a part of HST/SNAP 17110 (PI: D. Setton). The statistical sample program obtained imaging for a random subset of the 409 DESI Survey Validation LRGs with equivalent width $\rm H_{\delta,a}>4 \ \AA$, empirically selecting for all galaxies which may show signs of A-star or early B-star dominated spectra \citep[e.g.,][]{Balogh.etal.1999, French.etal.2015,Wu.etal.2018}. F110W was selected as the imaging band to maximize signal-to-noise in the 23.3 minute exposures while also observing at red enough wavelengths ($\lambda_\mathrm{rest}\sim5000 \rm \AA$) to measure light-weighted sizes that are reasonable proxies for the stellar mass-weighted sizes. In total, 182 galaxies in our sample were observed, however, the telescope lost locking on the acquired guide star and drifted during the exposure for 17 galaxies, resulting in artifacts in their final images. We remove 9 objects in the sample that have these artifacts in all four pre-drizzled single-shot images. For the remaining affected objects, we recombine the pre-drizzled single-shot images that are unaffected and perform analysis on the reprocessed image\footnote{Two additional objects are removed from this sample due to their spectra being low signal-to-noise and we cannot infer their star formation histories properly.}.

We obtained images using a 4-point dither pattern using the WFC3-IR-DITHER-BOX-MIN pattern to achieve optimal PSF-sampling, and observed using the MULTIACCUM mode with NSAMP=12 and SAMP-SEQ=STEP50, with 349.2 seconds/integration for a total integration time of 1396.9 seconds. The spatial resolution of these images is around $0.''13$, which translates to $\sim 1 \,kpc$ at $ z\sim 1$. In Appendix \ref{appendix: all_gallery}, we present a selected gallery of image cutouts, demonstrating the combination of sensitivity and resolution across the full range of stellar masses in our sample.

The PSFs on the HST/WFC3 IR detector are well characterized and stable, but vary spatially across the detector.  This spatial variation is described by a 3x3 array of empirically constructed PSF models in \cite{Anderson.2016}. Since the target galaxies are always centered in the imaging, we choose the center PSF in the 3x3 array to best represent the PSF of the target galaxy in each single-shot image. As these single-shot images are processed through \texttt{AstroDrizzle} \citep{AstroDrizzle} to make the final image products on which we perform the galaxy structure fitting, the empirically constructed PSFs are further distorted by the drizzle pattern in the image pipeline. To account for such distortion, we inject the center empirical PSF stamp into each pre-drizzled single-shot image such that the target galaxy is replaced by the empirical PSF. We then repeat the drizzling processes on these PSF-injected images, using the same \texttt{AstroDrizzle} inputs as documented in the drizzle log files. We finally extract the PSFs, which have captured the drizzle patterns, from the re-drizzled final outputs and use them for the aforementioned structure fitting analysis.

\subsection{3D-HST Comparison Sample}\label{3D-HST Comparison}
We choose the galaxies in the 3D-HST survey \citep{Brammer.etal.2012} to contextualize the size and structures of PSBs in this sample relative to the coeval star-forming and quiescent galaxies, given the availability of its auxiliary morphology catalog based on HST imaging data. We select only the 3D-HST galaxies that are within the same stellar mass ($\rm log(M{*}/M_{\odot}) > 10.65$) and redshift ($1<z<1.3$) ranges as the PSBs in this sample, using the stellar mass derived with \texttt{Prospector} \citep{Leja.etal.2019} and photometric redshifts in \cite{Skelton.etal.2014}. The photometric redshifts of 3D-HST galaxies are derived with multiple photometric bands and are sufficiently accurate, with a normalized median absolute deviation $<2.7\%$ when compared to the available spectroscopic redshifts in the fields. Since the angular diameter distance does not vary drastically at these redshifts ($1<z<1.3$), the systematic errors in derived physical sizes are small and the comparison of physical sizes is not affected significantly by the difference between photometric and spectroscopic redshifts. We include their best-fitting S\'ersic parameters measured in the F125W filter from the catalog of \cite{vanderWel.etal.2014}, which are the closest to the wavelength range of the F110W filter among the available catalog data. We further omit the galaxies flagged as bad fits. We note that although these best-fitting parameters were measured using a different code (\texttt{GALFIT}), we find no systematic biases between fits done with Pysersic and Galfit, see details in Section \ref{sec:size measurement} and Appendix \ref{fig:Appendix_B}.
We separate star-forming and quiescent galaxies in the 3D-HST sample with the UVJ rest-frame colors derived in \cite{Skelton.etal.2014} and the color cut in \cite{vanDokkum.etal.2015}.

The post-starburst galaxies in this sample have high stellar masses ($\rm 10.65 < log(M_*/M_{\odot}) < 11.8 $), a regime where the number densities for PSBs, quiescent and star-forming galaxies are low \citep{Wild.etal.2016}. Due to the size limitation of the 3D-HST survey footprint and the relative scarcity of objects in this mass range, we lack the number of high stellar mass quiescent galaxies or star-forming galaxies to assemble a mass-matched coeval comparison group for the PSBs in this sample. We instead choose to extrapolate the properties of the quiescent and star-forming populations as a smooth function of stellar mass in the subsequent analysis. We divide the quiescent galaxies and star-forming galaxies into six mass bins over the range $\rm log(M_{*}/M_{\odot}): 10.65-11.8$, with each bin to be roughly $\rm \sim 0.2 \, dex$ wide. We calculate the median sizes of star-forming galaxies and quiescent galaxies in each bin and the corresponding median errors via the bootstrapping method implemented by the standard routine in \texttt{Scipy} \citep{2020SciPy}. We fit a power-law scaling relation similar to the one in \cite{vanderWel.etal.2014} to the median stellar mass and median sizes:
\begin{equation}\label{eq:powerlaw}
     R_{e}/(kpc) = A(\frac{M_{*}}{10^{11.1}M_{\odot}})^{\alpha},
\end{equation}
where the best-fitting slope $\alpha$ and normalization $A$ are determined via Markov Chain Monte Carlo (MCMC) sampling implemented by \texttt{emcee} \citep{emcee}, assuming a likelihood function:
\begin{equation}
\begin{split}
     \ln \,p(R_{e}|M_{*},\sigma,A,\alpha) = \\
     -\frac{1}{2} \sum( \frac{(R_{e}/(kpc)- R(M_{*},A,\alpha))^2}{\sigma^2}),
\end{split}
\end{equation}
where $R(M_{*},A,\alpha)$ is Equation \ref{eq:powerlaw} and $\sigma$ are the errors on median $R_{e}$. In the following figures, we visualize the errors in our fitted relation by drawing the 16th and 84th percentile of the best-fit $A$ and $\alpha$ parameter posterior distributions and plotting the corresponding scaling relationship. We extrapolate the mass-central density relation similarly for quiescent galaxies and star-forming galaxies.

\section{Structure Fitting} \label{sec:method}
\subsection{Size Measurements} \label{sec:size measurement}

 \begin{figure*}[!htb]
    \centering
    \includegraphics[width = \textwidth]{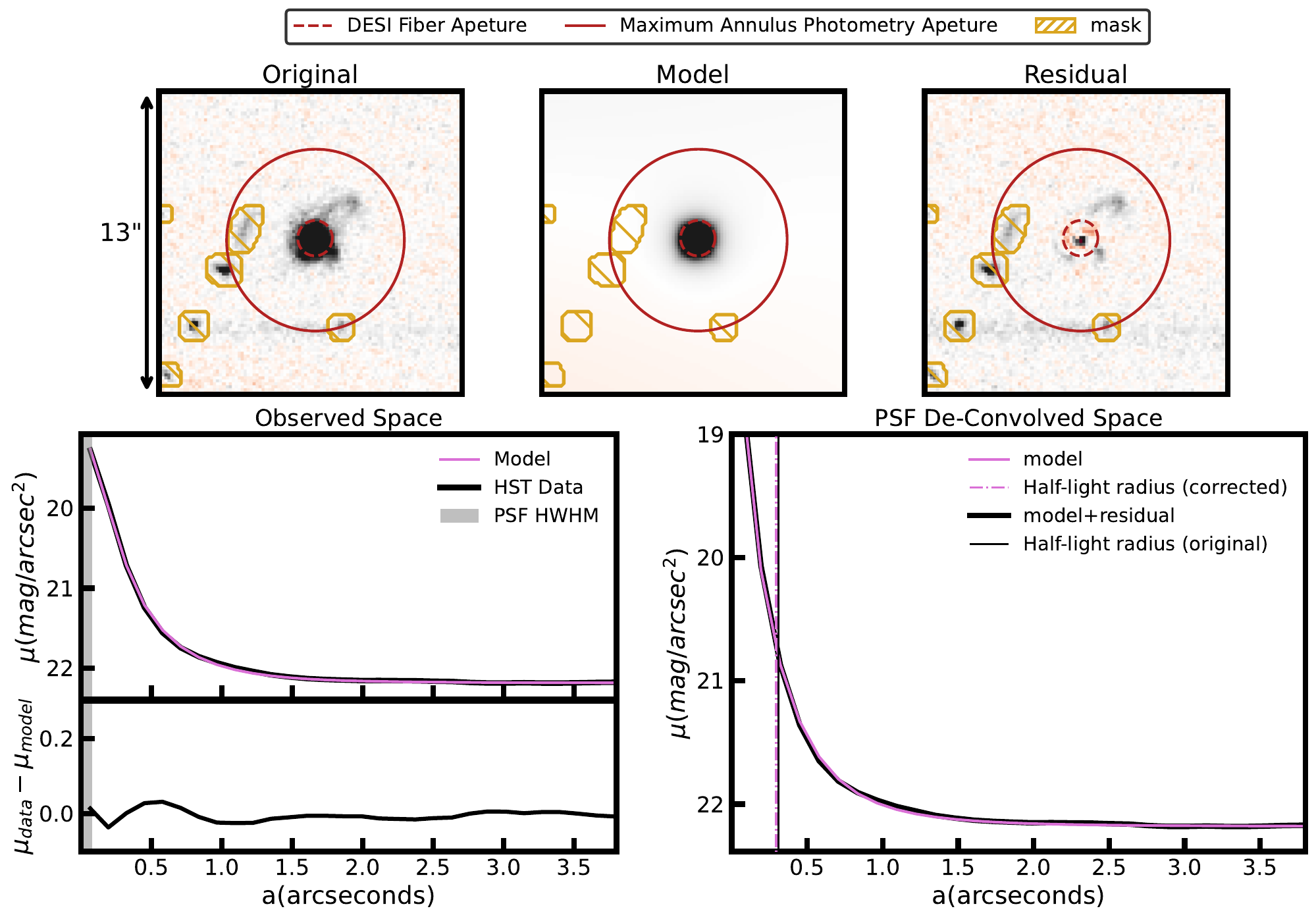}
    \caption{(Upper left) Original image, (Upper middle) best-fitting 2D S\'ersic model, and (Upper right) residual of an example post-starburst galaxy with clear tidal features. The color map is chosen so that positive pixel values are black, negative pixel values are red, and zero pixel values are white. The red solid ellipse in each upper panel traces the largest annulus used for profile extraction. The red dashed circle in each upper panel traces the aperture size of DESI fiber centered on the pointing of this target. The yellow polygons with hatches in each upper panel mark the regions in the image data that are masked out during the model fitting. (Lower left) Radial surface brightness profiles as a function of the semi-major axis of the best-fitting ellipse of original data (Top; black), best-fitting model (Top; pink), and residual (Bottom; black) extracted with annular apertures centered on the post-starburst galaxy. (Lower right) Radial surface brightness profiles of the best-fitting S\'ersic model (pink) and the same model with the residual added (black), using the method in \cite{Szomoru.etal.2012}. There is minimal difference between the S\'ersic half-light radius (black solid line) and the residual-corrected value. }
    \label{fig:method}
\end{figure*}

We use \texttt{pysersic} \citep{pysersic}, a Bayesian structural fitting tool, to fit S\'ersic profiles to the galaxies in this sample. We use \texttt{SEP} \citep{Barbary2016,Bertin.etal.1996} to identify and deblend sources in each image and generate corresponding segmentation maps. Based on the segmentation maps, we create masks for sources that are either 2 mags dimmer than the target galaxy or visually identified as a point source. We enlarge the masks by 3 pixels through kernel convolution in standard \texttt{SciPy} routines, in order to fully mask light from all interloping sources. We assume a single S\'ersic profile for each unmasked source and a flat sky background, which we fit simultaneously. 

We perform structural fitting with the same single S\'ersic profile setup using \texttt{GALFIT} \citep{GALFIT} as a cross-check to the measurement we obtained with \texttt{pysersic}. We find no significant systematic biases in the measurements of these two fitting routines. ($ \rm median (log(R_{e, pysersic})-log(R_{e, GALFIT})) \sim 0.03 \, dex$) A detailed comparison of results from the two fitting routine is included in Appendix \ref{appendix: method_comparison}. In the subsequent analysis of this paper, we choose to report the PSB morphological properties measured by \texttt{pysersic}. 

We measure the fraction of flux enclosed in the center region (an ellipse with the best-fitting axis-ratio and a semi-major axis of $\rm 1kpc$) for all galaxies using their best-fitting S\'ersic profiles. Assuming a uniform mass-to-light ratio, we compute the fraction of stellar mass contained in the same area. Then we derive the stellar mass surface density within one kiloparsec ($\Sigma_{\rm 1kpc}$) for these galaxies: 
\begin{equation}
    \Sigma_{\rm 1kpc} = \frac{M_{*}\frac{f_{\rm 1kpc}}{f_{\rm tot}}}{\pi(\rm 1 kpc)^2(b/a)}.
\end{equation}

\subsection{Accounting for Deviations from S\'ersic Profiles}
We find symmetric low surface brightness features (eg., blobs and strips) close to the target galaxies in several cases, through visual inspection of image cutout and modeling residual, see Appendix \ref{appendix: all_gallery} for examples. The presence of these features in some galaxies suggests that they are tidally disrupted by recent merger events or have nearby companions. Given that these faint features are close to or completely blended in the target galaxies, we could not properly extract them in the segmentation map or mask them during the modeling process. As a result, the fitting routine can be driven to compensate for the additional flux brought by these features in the galaxy outskirts, biasing the half-light radius and/or the S\'ersic index. In addition, the morphology of post-starburst galaxies may not necessarily follow a perfect single S\'ersic profile. The systematic choice of using a single profile in modeling can result in bias in the subsequent analysis.

We follow the procedure in \cite{Szomoru.etal.2012} to account for the impact of any non-single-S\'ersic flux components in our size measurements. After running the structural fitting routines described in Section \ref{sec:size measurement}, we extract the model residual and apply the same mask used in the fitting to the residual for each instance. We then perform annular photometry on the masked residual, using the best-fitting x-y position, angular position, and axis ratio from each 2d S\'ersic model, to extract the radial profile of residual flux. We add this residual profile (PSF-convolved) back into the analytical 1D model profile (non-PSF-convolved). Finally, we measure the half-light radius of the combined profile in 1D space through numerical integration.

In general, single S\'ersic profiles accurately capture the 2D light distribution of the target galaxies in this sample, even for those that are tidally disrupted. In Figure \ref{fig:method}, we show an example fit for a galaxy with obvious tidal disruption features. For this object, the radial surface brightness residuals are minimal ($\rm max \{\mu_{obs} - \mu_{model}\} < 0.2 \, mags/arcseconds^2$) and the corrected $R_{e}$ values do not deviate significantly from the $R_{e}$ values measured directly ($\rm |R_{e-corrected} - R_{e-direct} | \sim 0.04 \, arcseconds$). A complete evaluation of the impact of these residual corrections is available in Appendix \ref{appendix: method_comparison}.

\subsection{Close Pairs}
Five objects in this sample resolve into close pairs of galaxies in the HST imaging. In all cases, the DESI fibers enclose light from both components. We find no evidence of multiple sets of spectral features occurring at different wavelengths in the spectra of these pairs, suggesting that the enclosed components in each pair have similar redshifts or drastically different redshifts. We assume these pairs are interacting systems in the following analysis, though we cannot fully rule out the possibility that these pairs are superpositions of isolated objects. We note that the stellar population properties and star formation histories derived from SED fitting with \texttt{Prospector} in these cases are composite results of both components. We are unable to specify the stellar population properties of each component or to determine whether both components are ``Post-Starburst". 

In the aforementioned structure fitting analysis, we model each component separately as a single S\'ersic profile. We find that in all cases, the components in each pair have similar magnitudes (differences are within 1 mag). We choose to report the fitting results of the component that is closest to the fiber-pointing for each pair. We expect the reported sizes of these systems to be underestimated and not reflect their true location on the mass-size plane, since we report the stellar mass that is the sum of each pair. Throughout this paper, we denote close pairs with different symbols and derive scaling relations without that subsample.

\section{Results}\label{sec: analysis}
\subsection{PSBs Exhibit Compact Sizes} \label{sec: mass_size}

\begin{figure}[!htb]
    \centering
    \includegraphics[width = \linewidth]{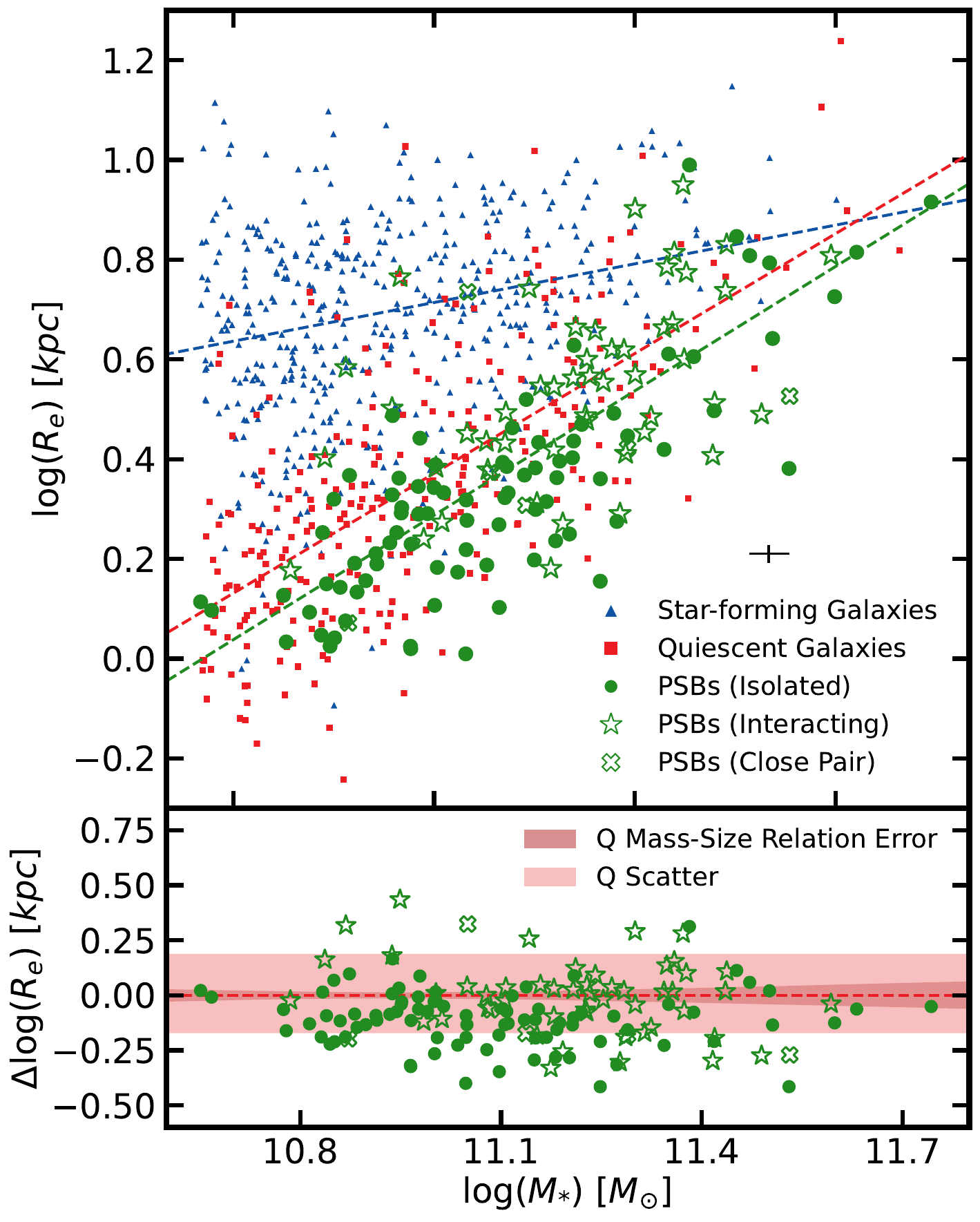}
    \caption{Upper panel: The semi-major effective radius versus stellar mass relation at $1<z<1.3$ for post-starburst galaxies (green, this sample), star-forming (blue) and quiescent (red) 3D-HST galaxies. Symbol shapes indicate isolated (circles) and interacting (stars) PSBs. The best-fitting relations for each population are shown as dashed lines and typical error bars are indicated in the lower right corner. Lower panel: Deviations from the quiescent size-mass relation. The error on the best-fitting quiescent relation is shown as a shaded region in dark red and the scatter of quiescent galaxies around the best-fitting quiescent relation is in light red. On average, the post-starburst galaxies in this sample are more compact than both star-forming and quiescent galaxies. }
    \label{fig:mass_size}
\end{figure}

On the mass-size ($M_*$-$R_{e}$) plane, the galaxy sizes scale with their masses and different galaxy populations display distinct scaling relations across different epochs. The quiescent mass-size relation is generally steeper and has a lower normalization than that of star-forming galaxies \citep{vanderWel.etal.2014,Whitaker.etal.2017, Wu.etal.2018,Mowla.etal.2019,Martorano.etal.2024}, reflecting the more compact morphology of quiescent galaxies. By comparing the sizes ($R_{e}$) of PSBs with their star-forming and quiescent counterparts from the same epoch, we aim to empirically constrain the evolutionary pathways of PSBs and shed light on their quenching modes.

In the upper panel of Figure \ref{fig:mass_size}, we show the effective radius versus stellar mass for the galaxies in this sample in green. The 3D-HST star-forming and quiescent galaxies at ($1<z<1.3$), for which sizes are measured in the HST/F125W imaging, are plotted in blue and red respectively in the same panel \citep{vanderWel.etal.2014}. We fit power-law mass-size relations for star-forming galaxies, quiescent galaxies, and PSBs following the procedure described in Section \ref{3D-HST Comparison} and show these relations as blue, red, and green dashed lines respectively. In addition, we mark the galaxies identified as having clear tidal or merger features during the visual inspection as green star symbols. The masses used here for star-forming and quiescent galaxy populations are derived through \texttt{Prospector} while assuming a non-parametric SFH \citep{Leja.etal.2019} and have median values about 0.2 dex higher than those in \cite{vanderWel.etal.2014}. 

We confirm that these massive PSBs have much smaller half-light radii than star-forming galaxies and have sizes similar to, but even smaller than, the quiescent population, as found in several previous studies \citep[e.g.,][]{Yano.etal.2016,Maltby.etal.2018,Wu.etal.2018,Wu.etal.2020,Setton.etal.2022,Clausen.etal.2024}. However, we do note that this comparison is limited by the small number of star-forming galaxies and quiescent galaxies available in the most massive regime ($\rm log(M_{*}/M_{\odot}) > 11.3$) and the extrapolation becomes less reliable. To probe the subtle difference in size between quiescent galaxies and PSBs, we measure the vertical offset of PSBs from the quiescent mass–size relation, same as equation 2 in \cite{Setton.etal.2022}:
\begin{equation}
\rm \Delta log(R_{e}) = log(R_{e}) - log(\hat{r}_{e},Q(log\frac{M_{*}}{M_{\odot}})).
\end{equation}

In the lower panel of Figure \ref{fig:mass_size}, we show $\rm log(M_{*})$ versus $\rm \Delta log(R_{e})$ for these PSBs. Overall, the PSBs in this sample are slightly more compact than their contemporary quiescent population, lying systematically $\rm \sim 0.075 \, dex$ below the quiescent mass-size relation while the scatter in sizes is $\rm \sim 0.1 \, dex$. If the same analysis is performed with PSB sizes measured with GALFIT, the PSBs then lie systematically $\rm \sim 0.08 \, dex$ below the quiescent mass-size relation. To robustly test the statistical significance of such an offset, we choose to perform a two-sample Kolmogorov–Smirnov test on the PSB and quiescent distribution in $\rm \Delta log(R_{e})$, using standard implementation of this method in \texttt{Scipy}. We obtain a p-value of 0.00036, which rejects the null hypothesis that PSBs and quiescent galaxies are drawn from the same distribution in relative sizes. Such exceptional compactness is consistent with PSBs at slightly lower stellar masses ($ \rm 10 < log(M_{*}/M_{\odot}) < 11.2$) at similar redshifts \citep{ Almaini.etal.2017} and PSBs at lower redshifts ($\rm z \sim 0.7$) with similar stellar masses \citep{Setton.etal.2022}. 

\subsection{Dense Cores within Post-Starburst Galaxies}\label{sec: mass_sigma1}

\begin{figure}[!tb]
    \centering
    \includegraphics[width = \linewidth]{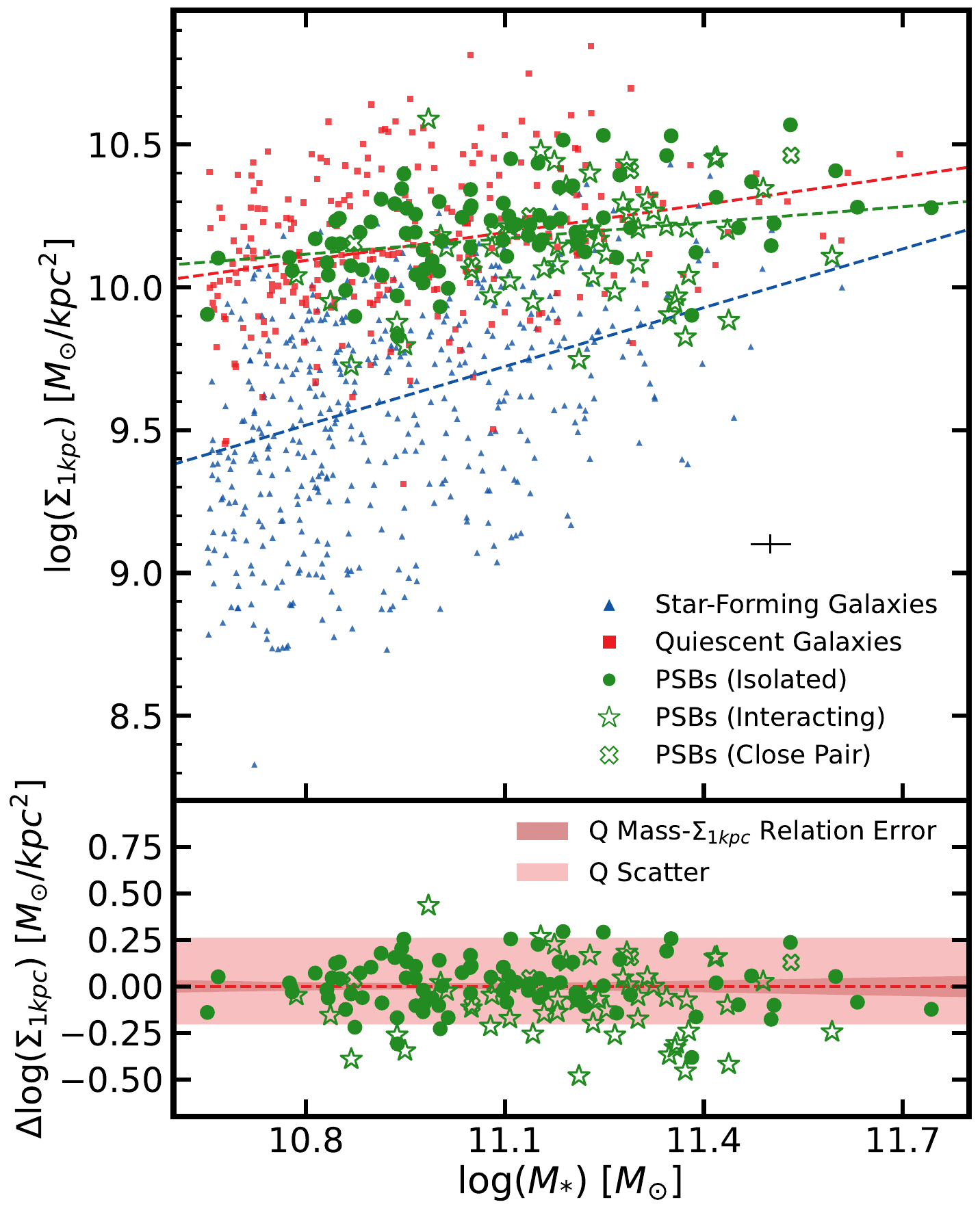}
    \caption{The surface mass density in the central one kiloparsec versus stellar mass relation for post-starburst galaxies (green), star-forming (blue) and quiescent (red) 3D-HST galaxies (upper panel) and residuals from the quiescent relation (lower panel). Plotting conventions as in Figure \ref{fig:mass_size}. Despite their more compact sizes, the post-starburst galaxies in this sample have similar central surface mass densities as the quiescent galaxies. }
    \label{fig:mass_surfacedensity}
\end{figure}

Similar to half-light radii, the central densities ($\rm \Sigma_{1kpc}$) of galaxies also scale with masses \citep[e.g.,][]{Fang.etal.2013,Barro.etal.2017,Mosleh.etal.2017}. Typically, the star-forming relation has a lower normalization but a steeper slope than the quiescent relation, reflecting the growth of the bulge during the star-forming phase. In the upper panel of Figure \ref{fig:mass_surfacedensity}, we present the central surface mass densities ($\rm \Sigma_{1kpc}$) versus stellar mass for PSBs (greens), star-forming galaxies (blue), and quiescent galaxies (red). The extrapolated star-forming, quiescent, and post-starburst relations are plotted as blue, red, and green dashed lines respectively. The colors and marker signs of PSBs are similar to those in Figure \ref{fig:mass_size}. In the lower panel, we show the distribution of relative central surface mass densities ($\rm \Delta log(\Sigma_{1kpc})$) versus stellar mass. This quantity is defined as the vertical offset of PSBs to the quiescent relation in the $\log\Sigma_{1kpc}-M_{*}$ plane, similar to $\rm \Delta log(R_{e})$.

The PSBs follow a relatively shallow $M_{*} - \Sigma_{1kpc}$ scaling relationship and have a median $\rm log(\Sigma_{1kpc}) \sim 10.1 $. Overall, the median value and scaling relation of PSB central densities are similar to those of quiescent galaxies in this comparison sample($\rm \Delta \Sigma_{1kpc, PSB-Quiescent} \sim - 0.07 \, dex$), and to those of quiescent galaxies found in the literature \citep{Barro.etal.2017,Mosleh.etal.2017}. Furthermore, these PSBs have much denser inner structures than the star-forming galaxies around the same mass ($\rm \Delta \Sigma_{1kpc, PSB-SFG} \sim  0.5 \, dex$). Such similarity between PSBs and quiescent galaxies at $z\sim 1.1$ is consistent with the PSBs above $z\sim 0.7$ studied in \cite{Setton.etal.2022} and indicates that the structural differences between the two populations are due to the more extended outskirts of the older quiescent population.

\begin{figure*}[t]
    \centering
    \includegraphics[width = \textwidth]{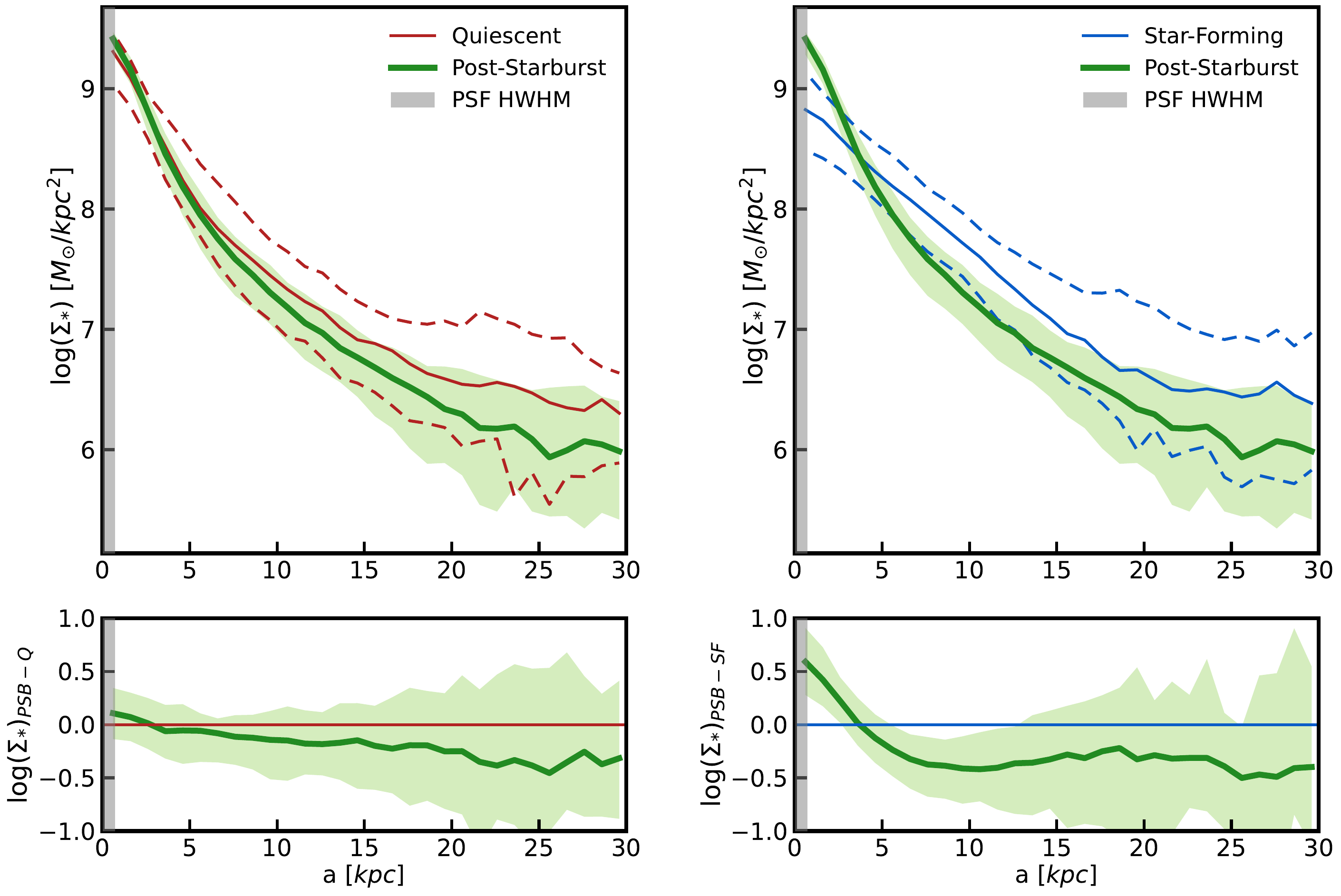}
    \caption{Upper panels: median surface mass density profiles along the semi-major axes for ($\rm log(M_{*}/M_{\odot}) < 11.2 $) PSBs (black) versus mass-matched quiescent (left) and star-forming (right) galaxies. The PSB population scatter is shown as a shaded region in light green and scatter amongst quiescent and star-forming galaxies is indicated by dashed lines. The half-width at half maximum of the HST/F110W PSF is shown as a gray band. The impact of PSF is negligible beyond $a>1\,\rm kpc$ . Lower panels:  Residuals of subtracting the median quiescent (left) and star-forming (right) median profiles from the median PSB profile. The intrinsic scatter of the differences between PSB and quiescent/star-forming galaxies is shown as a shaded region in light green in each panel respectively.} 
    \label{fig:stacked_profile}
\end{figure*}

The mass profiles derived in observed space capture and demonstrate any potential deviations from the S\'ersic profiles. In the upper panels of Figure \ref{fig:stacked_profile}, we show the light-weighted median mass surface density profiles of quiescent galaxies, star-forming galaxies, and PSBs in red, blue, and black solid lines and mark the scatter of their profiles with red, blue dashed lines, and a gray-shaded region. In the lower panels, we show the residual profiles in black and their scatters in gray by subtracting the median quiescent profile or the median star-forming profile from the median PSB profile. To obtain these light-weighted median mass density profiles, we first measure the empirical surface brightness profile with annular photometry. Assuming the mass-to-light ratio is the same along the profile, we then derive the mass-to-light ratio individually for each galaxy by taking the SED-inferred stellar mass over the integrated flux using the brightness profile. Finally, we renormalize the surface brightness profile of each galaxy by the corresponding mass-to-light ratio. For each stacked PSB, we match and stack a quiescent as well as a star-forming 3D-HST galaxy that has similar stellar mass ($\rm |log(M_{*, PSB}) - log(M_{*, Quiescent/Star-forming})| < 0.1\,dex$). Since we do not have enough 3D-HST galaxies on the high mass end for matching, we only choose galaxies at $\rm log(M_{*}/M_{\odot}) < 11.2 $ and match 60 sets of galaxies in total. We note that the profiles in this analysis are derived from raw observational data and therefore include the PSF. The overall flux distribution is altered by the PSF and the subsequently derived mass density profiles have lower numerical values at $a = 1\,\rm kpc$ than the $\Sigma_{1kpc}$ values reported in previous sections, which are derived in non-PSF-convolved space. All the star-forming galaxies and quiescent galaxies matched are from the same redshift interval $1<z<1.3$ as this sample. The angular diameter distance does not vary significantly within this redshift interval and the galaxies stacked in this analysis have similar PSF HWHM effective sizes. We overplot the half width at half maximum (HWHM) of the WFC3/F110W imaging PSF, which we derived using the angular diameter distance at the median redshift of this sample ($z=1.15$), as a gray-shaded region in each panel. The PSF HWHM of F125W imaging is similar to that of F110W. We deem the comparison of the overall mass profile shape to be sufficiently robust for $a > 1\, \rm kpc$ for these different galaxy populations.

The comparison of these profiles validates the previous conclusion that both the typical quiescent galaxies and PSBs have much smaller half-light radii and higher central densities than the typical star-forming galaxies. The systematic difference in the light-weighted mass profiles further reveals what is driving the fits of PSBs to have overall smaller sizes than those of quiescent galaxies, despite the fits of these two populations field similar central densities.

\subsection{Other parametric measures of structure}

\begin{figure*}[!t]
    \centering
    \includegraphics[width = \textwidth]{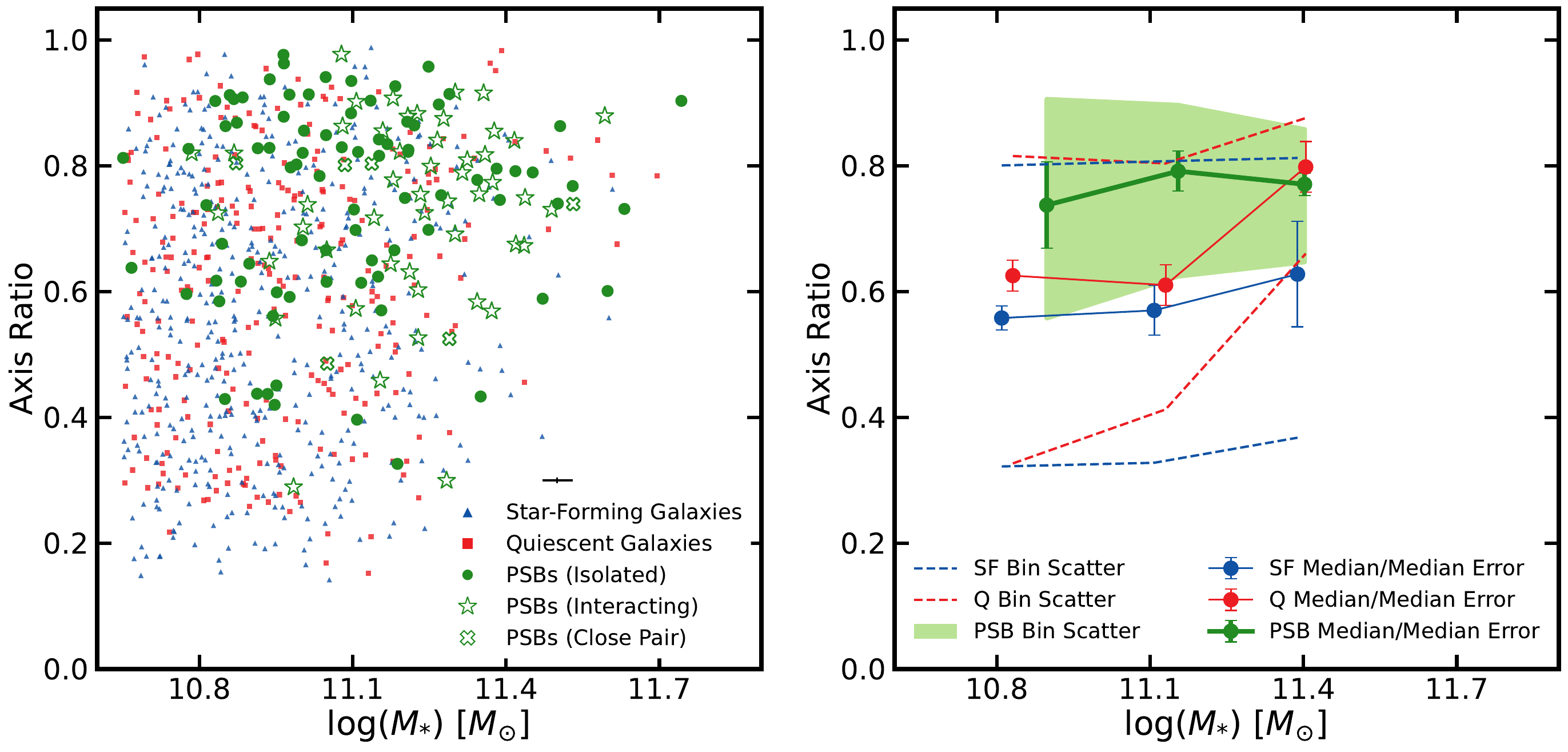}
    \caption{Left panel: Axis-ratio versus stellar mass for PSBs (green), 3D-HST quiescent galaxies (red), and star-forming galaxies (blue). The star symbols mark the PSBs with merger features (interacting) while the circle symbols mark the ones that are not disrupted (isolated). Right panel: Running median and 16-84th population distributions for each galaxy population. With some notable exceptions, the PSBs in this sample are round in projection, suggesting that they are predominantly round spheroids.}
    \label{fig:axis_ratio}
\end{figure*}

Parameters used in our structure fitting other than half-light radius such as axis-ratio and S\'ersic index give us additional insights into the 3D structures of galaxy populations. In Table \ref{tab:all_measurements}, we tabulate the results and uncertainties of measured axis-ratios, S\'ersic indices, and half-light radii as well as derived $\rm \Sigma_{1\,kpc}$ of all PSBs. In the left panel of Figure \ref{fig:axis_ratio}, we show the measured projected axis-ratios as a function of stellar mass for Star-forming galaxies (blue), quiescent galaxies (red), and PSBs (green). In the right panel, we over-plot the mass-binned ($\rm log(M_{*}/M_{\odot}): [10.6-11],[11-11.3],$ and $[11.3-11.8]$) medians (filled circles) and population scatter(filled band for the PSB sample, dashed lines for 3D-HST comparison). 

Although incredibly elongated examples exist, the PSBs predominantly have large projected axis ratios (median $\rm b/a \sim 0.8$; nearly circular in projection) in all mass bins, which is consistent with the finding in \cite{Clausen.etal.2024}. We perform axis ratio modeling to infer the intrinsic 3D shapes of these galaxies, following a similar methodology to \cite{Letter.vanderWel.etal.2014} modified to use a Bayesian framework (using the code the \texttt{BEAST};  for more details see \cite{Gibson.etal.2024} and S. Price et al., in prep.). The modeling suggests that nearly all ($\gtrsim90\%$) of these objects are round spheroids rather than disks, in contrast with the wide spread exhibited by randomly-oriented disks amongst the SFG population. Furthermore, given that PSB galaxies differ structurally from both comparison samples, their shapes must either change in transition or the channel only impacts a biased subset of the population. The prevalent large axis-ratios in our fits are not driven by the presence of tidal features; there is no significant distinction in the axis-ratio distribution between isolated and merging PSBs. Overall, our finding of these PSBs being compact spheroids is consistent with previous studies such as \cite{Maltby.etal.2018} and \cite{Almaini.etal.2017}. 

A number of quiescent galaxies in the comparison sample are elongated in projection ($b/a <0.5$), suggesting that these galaxies could have experienced color transformation without transforming into spheroids and bias the comparison between quiescent galaxies and PSBs. We note that these galaxies lie systematically $\sim 0.07 \,\rm dex$ below the quiescent mass-size relation with a scatter similar to the overall quiescent population. The $\Sigma_{1kpc}$ of these galaxies are $\sim 0.2 \,\rm dex$ above the median of the entire quiescent galaxy population at fixed masses. However, when removing this population of elongated quiescent galaxies in fitting, the quiescent relation in $M_{*} - \Sigma_{1kpc}$ is only lowered by $\sim 0.1 \,\rm dex$ in the low mass regime ($\rm log(M_{*}/M_{\odot}) < 11.2$) and the relation hardly changes at higher masses. Hence, our qualitative conclusions in Section \ref{sec: mass_size} and Section \ref{sec: mass_sigma1} would remain similar when considering the bias caused by these galaxies. 

The PSBs in this sample are fitted with relatively high S\'ersic indices ($\rm n_{median} \sim 3$), similar to quiescent galaxies. However, we do note that the numerical values of S\'ersic index can be sensitive to the exact implementation of fitting routine, in the high sersic index regime ($n > 2$). We choose not to make a quantitative comparison between the S\'ersic indices measured for quiescent galaxies (\texttt{GALFIT}) and PSBs (\texttt{pyseric}).

\begin{table*}[thbp]
\caption{Related properties of $\rm H_{\delta, a} > 4 \AA$ galaxies in this work, including DESI ID, redshift, stellar mass, star-formation rate, fraction of mass formed in the last one billion year, time since quenching, half-light radius, S\'ersic index, axis-ratio, and central surface mass density. Part of the table is truncated for visualization purposes.}
    \centering
    \begin{tabular}{c|cccccccc}
    \hline
    \hline
      DESI ID & $z$& $\rm log(M_{*})$ & $SFR$ & .. & $\rm R_{e} $  & n  & b/a  & $\rm log(\Sigma_{1kpc})$  \\
       & & [$\rm h^{-2} M_{\odot}$]& $\rm [h^{-2} M_{\odot}yr^{-1}]$ & & $\rm [h^{-1}kpc]$ & & &[$\rm M_{\odot}/kpc^2$] \\
     \hline
     39627196272216758  &1.1255& $10.94^{+0.04}_{-0.04}$&
      $0.42^{+0.63}_{-0.4}$& .. & $1.79^{+0.02}_{-0.02}$ & $5.98^{+0.01}_{-0.03}$& $0.561^{+0.005}_{-0.006}$&  $10.35^{+0.05}_{-0.04}$\\
     39632965117937224  &1.2215& $11.10^{+0.03}_{-0.02}$&
      $1.92^{+0.89}_{-0.55}$& ..& $1.86^{+0.01}_{-0.01}$ & $2.43^{+0.04}_{-0.04}$& $0.884^{+0.005}_{-0.006}$&  $10.17^{+0.03}_{-0.03}$\\
     39628209112747541  &1.2645& $11.42^{+0.03}_{-0.03}$&
      $5.65^{+1.98}_{-2.36}$& ..& $3.26^{+0.02}_{-0.03}$ & $3.78^{+0.04}_{-0.04}$& $0.675^{+0.003}_{-0.003}$&  $10.46^{+0.03}_{-0.03}$\\
     39633047741531159  &1.0227& $11.63^{+0.03}_{-0.02}$&
      $0.03^{+0.08}_{-0.02}$& ..& $6.53^{+0.07}_{-0.06}$ & $3.36^{+0.03}_{-0.03}$& $0.732^{+0.003}_{-0.003}$&  $10.28^{+0.03}_{-0.02}$\\
      .. &  .. & .. & .. & .. & .. & .. & .. & .. \\
      \hline
    \end{tabular}
    {\footnotesize \tablecomments{This table will be available in its entirety in a machine-readable form.}}
    \label{tab:all_measurements}
\end{table*}

\section{Discussion}\label{sec: discussion}
\subsection{Looking backward: How was the compact structure formed?}

\begin{figure*}[!htb]
    \centering
    \includegraphics[width = 0.9\textwidth]{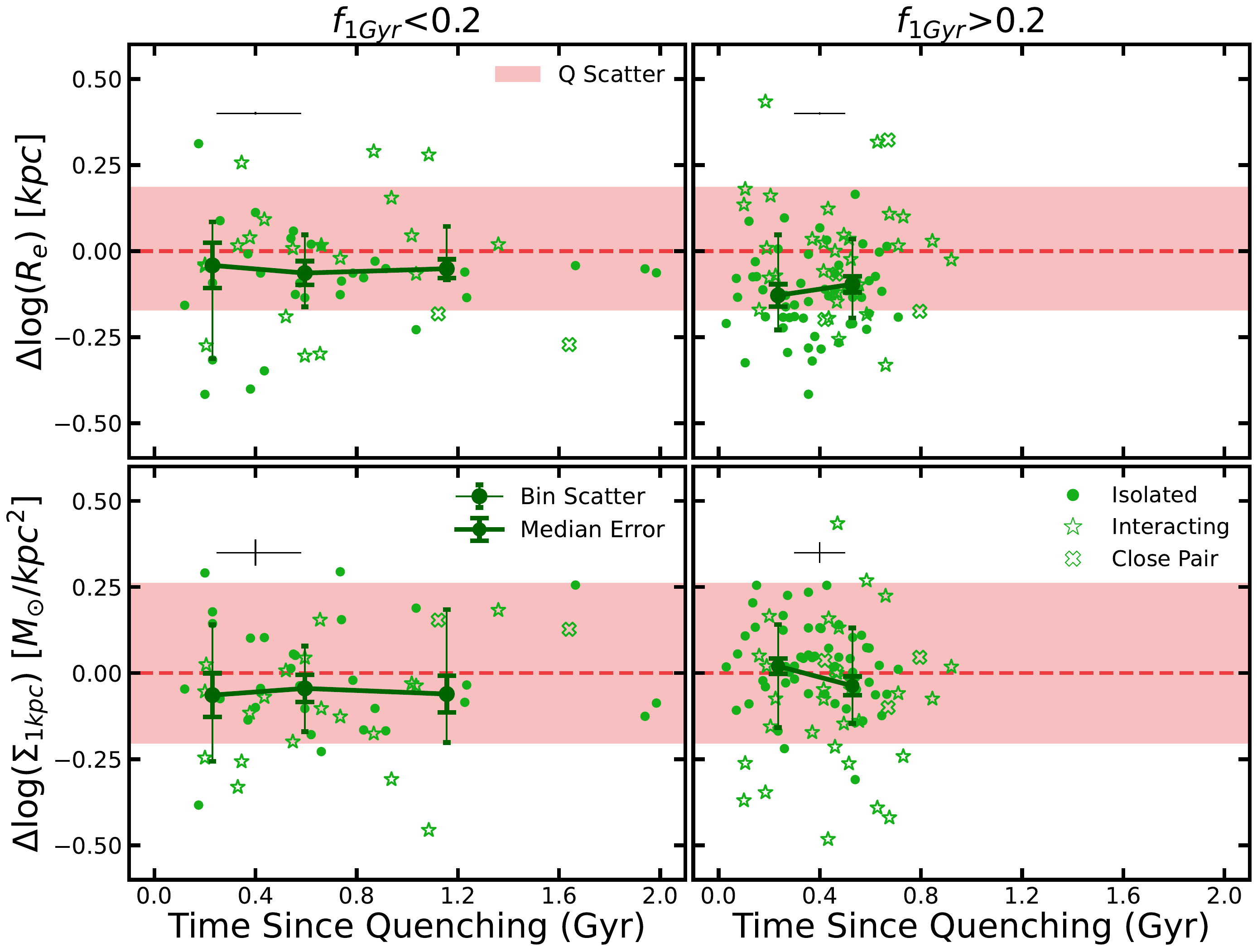}
    \caption{PSB sizes (top) and central densities (bottom) relative to quiescent galaxies versus time since quenching as defined in Section \ref{sec:PSB Selection}. PSBs are subdivided into with small ($f_{1Gyr}<0.2$, left) and significant bursts ($f_{1Gyr}>0.2$, right). The PSB median properties, errors on the medians (thick error bars), and population scatter (thick error bars) in each $t_{q}$ bin are shown in dark green. Typical error bars are included in upper left corners. The scatter of quiescent galaxies around the best-fitting $\rm M_*$-$R_{e}$ or $\rm M_*$-$\Sigma_{1\,kpc}$ quiescent relation is shown as shaded regions in light red.}
    \label{fig:4panels}
\end{figure*}

If the progenitors of these PSBs are the typical star-forming galaxies with extended structures (represented by the star-forming relation in $\rm M_*$-$R_{e}$ in Figure \ref{fig:mass_size}), one likely pathway for them to achieve the compact structures seen in this sample is going through a central starburst phase. If a large fraction of stellar mass (the median $\rm f_{1Gyr}$ is around 0.4 in this sample) formed within a spatial scale much smaller than the original half-light radius, the centers of the galaxy light profiles would be skewed by the burst. Since the young stellar population is orders of magnitude brighter than the old, the half-light radius of the galaxy would shrink toward the half-light radius of the burst component, assuming a non-negligible burst fraction. The derived star formation histories suggest that the recent bursts also drastically increased stellar masses ($\sim  0.2\, \rm dex$, given a typical $\rm f_{1Gyr} \sim 0.4 $  and a typical $\rm log(M_{*}/M_{\odot}) \sim 11.2$ ). Thus, such a process could effectively move star-forming galaxies onto the PSB relation.

The formation of compact structures in galaxies due to a central starburst has been seen in several simulation works \citep{Zolotov.etal.2015, Tacchella.etal.2016}, which can be caused by disruption events due to gas instability in star-forming disks \citep{Dekel&Burkert.etal.2014}. In this case, the dissipative processes, which may be induced by a disruptive gas inflow or gas-rich mergers, cause the gas in galactic disks to lose angular momentum, fall toward the inner region of the galaxies, and then trigger a compact burst of star formation in the center. As the stellar surface density reaches a certain physical upper limit \citep[e.g.,][]{Franx.etal.2008} and the cool gas is depleted at the end of the starburst episode, these galaxies become quiescent if there is no further inflow of cool gas. This scenario is supported by several observational works focusing on a small sample of PSBs at $z \sim 0.8$ such as \cite{D'Eugenio.etal.2020b} and \cite{Wu.etal.2020}, which exhibit inverse age gradients or color gradients that require the superposition of a compact younger stellar population on top of a more extended old stellar population. The compact sizes of starburst components ($\sim 0.1\, \rm kpc$) observed in massive starburst galaxies further bolster this point of view \citep{Sell.etal.2014, Diamond-Stanic.etal.2021}.

Alternatively, the PSBs in this sample can be descendants of star-forming massive compact galaxies at higher redshifts, which are morphologically similar to the quiescent massive compact galaxies found at lower redshifts \citep{vanDokkum.etal.2015,Barro.etal.2017}. In this scenario, the compact structure is already in place for these galaxies while they are still forming stars. They retain roughly constant sizes and move along a horizontal track until reaching the preferential location in the mass-size plane where they shut off rapidly.

Although, the two physical scenarios would leave clear imprints on the age and therefore color gradients of post-starburst galaxies, their compact sizes and single-band HST imaging hinders such a test. However, the dominance of a central burst on the light profiles would fade on $\sim$ Gyr timescales, as the M/L ratio of the galaxies would rapidly evolve as the youngest stars die off. Thus, the post-starburst evolution of size and central density relative to their quiescent descendants can also discriminate between the two scenarios. In the case of the central starburst scenario, we expect to see a growth in relative size and a decrease in the relative central density as the burst ages. In some extreme cases, the half-light radius growth could be significant (change in $\rm log(R_{e}/kpc) \sim 0.2\,dex$) in $t_{q} < 1\, \rm Gyr$, if the physical scale of the central starburst is smaller than $\rm 1\,kpc$ and the burst mass is less than $20\%$ of the total formed mass \citep{Setton.etal.2022}. In contrast, if the star-forming progenitors were already compact, we expect no evolution in the relative size or central density in $t_{q} < 1\, \rm Gyr$. Enabled by the deep spectroscopic data of this sample, we can infer the time since the quenching of every object and statistically examine whether there is a trend in relative size or central density as a function of $t_{q}$.

In Figure \ref{fig:4panels}, we show the scatter in relative size and central density versus time since quenching ($t_q$) for PSBs after minor (left, $f_{1Gyr}: [0-0.2]$) and major (right, $f_{1Gyr}: [0.2-1.0]$) starbursts based on derived burst mass fractions. We note that $f_{1Gyr}$ is a proxy for the mass formed during the burst period in the SFH, assuming the onset of the starburst episode was within $ 1\, \rm Gyr$ in the lookback time of these galaxies. This value is likely much lower than the true fraction of mass formed during the burst for those with large $t_q$. We overplot the median values in each age bin ($t_{q}: [0-0.3],[0.3-0.7],$ and $[0.7-2]$) in red. Overall, the median relative size and median central density of these galaxies do not evolve significantly as their youngest, most massive stars fade in the first $ 1\, \rm Gyr$, following both major and minor starbursts. The lack of significant growth in size or a decrease in central density among these galaxies suggests that the recent burst of star formation likely occurred with a physical radius greater than $\rm 1\,kpc$. This further implies that the bursts occurred in already compact galaxies. Individual PSB can still be growing after quenching, given the scatter in these parameter spaces.  However, the scatter is likely dominated by the inherited heterogeneity of the PSB population rather than any evolutionary effect. They could have intrinsically different sizes and central densities when they were quenched and these differences persisted until they were observed. 

Although central starbursts within $\rm 1\,kpc$ are unlikely the dominant formation pathway for these PSBs, the physical mechanism that drives the rapid quenching of star formation remains elusive. Merger-induced starbursts and the subsequent depletion of cold gas may be able to quench these galaxies abruptly. Previous observational works have unveiled the existence of extreme gas outflows in massive compact starburst galaxies \citep{Rupke.etal.2019, Perrotta.etal.2023,Davis.etal.2023}, which can even persist in the post-starburst phase \citep{Tremonti.etal.2007} given that some gas contents can still be retained in the galaxy after rapidly quenching \citep{Suess.etal.2017,Bezanson.etal.2022}. These outflows are driven by stellar radiation pressure and ram pressure contributed by stellar winds from a compact starburst component ($\sim 0.1\, \rm kpc$) and can far exceed the central escape velocity in these galaxies \citep{Diamond-Stanic.etal.2012,Diamond-Stanic.etal.2021}. Such extreme stellar feedback is able to expel a critical fraction of the cold gas and quench these galaxies without necessarily invoking AGN feedback \citep{Diamond-Stanic.etal.2012,Sell.etal.2014}. But it is unclear whether the extended starburst preferred in this sample could achieve comparable SFR surface densities to power those outflows with sufficient radiation pressure and ram pressure. Additionally, merger-induced starburst can still further funnel gas into the inner region of the galaxies, trigger AGN activities, and quench the star formation with short dynamical times ($t_{dyn} \sim 100 \rm \,Myrs$; \citealp{Hopkins.etal.2008}). But it is not sufficient to simultaneously shut off star formation at all radii, especially in the largest ($R_{e} > 4\, \rm kpc$) galaxies. Given the redshift range of this sample ($1<z<1.3$), we lack the spectral coverage ($5800\, \rm\AA < \lambda_{obs} < 9824 \, \rm\AA$) of diagnostic emission line signatures for AGN activities, such as $\rm [O III]\, \lambda5007/H\beta$, and the Wide-field Infrared Survey Explorer photometry in our data is not sufficiently deep to identify the existence of AGNs. 

Only a fraction ($\sim 40\%$) of the galaxies are identified as having merger signatures and we find no change in the fraction of mergers with $t_q$, which suggests mergers are weakly correlated to quenching in these PSBs. Objects with potential faint tidal features are likely misidentified as nondisruptive due to a limited surface brightness sensitivity in these images ($\mu_{max} < 22 \rm \, mags/arcseconds^2$), leading to an underestimated merger fraction. Other mechanisms have been proposed to link quenching to the stabilization of gas disk, such as morphological quenching \citep{Martig.etal.2009}, in which the gas disk stabilizes once the galaxy acquires a spheroid-dominated morphology. These mechanisms may explain the globally simultaneous shut-off in these galaxies, but it is not clear whether they could introduce a sufficiently abrupt shutdown to produce the spectra seen in PSBs. Spatially resolved spectra are needed to probe the distribution of stellar ages at different radii in these galaxies. This kind of information will allow us to discover any spatial pattern associated with the rapid decrease of star formation or the lack thereof in future studies, shedding light on the exact mechanism that quenches these galaxies in future studies.

The massive $z \sim 1.1$ PSBs in this sample are similar to those at $z \sim 0.8$ in \cite{Setton.etal.2022}, who plotted PSB size against $t_{q}$ and did not find any significant changes in PSB sizes since they quenched. The most massive PSBs at $z \sim 1.2$ are likely quenched in a similar way as those at $z \sim 0.8$. But at the same time, the evolutionary constraints placed on these PSBs seem to raise tension against other works on PSBs at intermediate redshifts, such as \cite{D'Eugenio.etal.2020b} and \cite{Wu.etal.2018} at $z \sim 0.8$, who favor the central starburst scenario. We emphasize that the PSBs in this sample represent a rare population in terms of their stellar mass ($\rm log(M_{*}/M_{\odot})_{median} \sim 11.2)$, which is about 0.5 dex higher than those in \cite{D'Eugenio.etal.2020b} and \cite{Wu.etal.2018}. It is reasonable to speculate that these PSBs with contradictory age gradients are descendants of different progenitors, representing variants of the fast-quenching evolutionary pathways that dominate different mass regimes. 

One possible progenitor for the PSBs at higher mass are the compact star-forming galaxies known as ``Blue Nuggets" \citep{Barro.etal.2013,Barro.etal.2014a,Barro.etal.2014b,Williams.etal.2014,Barro.etal.2017}. The theory is these galaxies form massive but compact structures through the dissipative compaction of the gaseous disk, similar to the mechanism in a central starburst scenario \citep{Dekel&Burkert.etal.2014,Zolotov.etal.2015}. At $z \sim 1.6$ ($\sim 1\,\rm Gyr$ before the median redshift of this sample), the number density of compact star-forming galaxies is $\sim 10^{-4} \, \rm Mpc^{-3}$ \citep{Barro.etal.2017}, which is about an order of magnitude higher than the number density of PSBs with the same selection criterion as this sample at $z\sim 1.2$ (at least $10^{-5} \, \rm Mpc^{-3}$; \citealp{Setton.etal.2023}). If a subset of these high redshift ``Blue Nuggets" experiences an episode of starburst at all radii on top of the existing dense core instead of quenching immediately, they would eventually form galaxies that are consistent with PSBs in this sample. It is unclear whether such a drastic star-forming episode can persist, since the gas contents in these ``nuggets" are generally low \citep{Spilker.etal.2016}, especially in the center \citep{Spilker.etal.2019}. Given that these objects are already massive and likely have high halo mass, it's also difficult for cold inflow gas to reach the center of these galaxies. The high gas density at earlier cosmic time may enable sufficient gas inflows that continue to fuel the intense stellar assembly at all spatial scales of those nuggets, after their initial period of starburst. 

\subsection{Looking forward: Minor mergers as the dominant mass growth mode.}

Compared to contemporary quiescent galaxies at the same masses, the PSBs in this sample are smaller while having similar central densities, as shown in Section \ref{sec: analysis}. At lower redshifts ($z \sim 0.8$) where the descendants of these PSBs would reside, the quiescent $M_{*}-R_{e}$ relation has a higher normalization than that at $z\sim 1.1$ \citep{vanderWel.etal.2014} while the quiescent $M_{*}-\Sigma_{1kpc}$ relation hardly evolves from $z\sim 1.1$ to $z\sim 0.8$ \citep{Barro.etal.2017}. If these PSBs stay quenched and indeed evolve into typical quiescent galaxies at lower redshifts, they must increase their sizes while maintaining a constant core structure. Furthermore, we have already shown these PSBs lack any significant growth in relative sizes in the first $1 \rm Gyr$ since quenching as their recently formed stars fade. This suggests that the mechanism that gives these PSBs the typical quiescent structure is likely some form of mass growth in the outskirts, which operates at a longer timescale than $1 \rm Gyr$, rather than passively fading. 

In Figure \ref{fig:stacked_profile}, the mass-matched quiescent galaxies appear to have higher median stellar mass densities than the contemporary PSBs starting from $a = 6\, kpc$, and the differences between their light-weighted mass density profiles increase at larger radii. In comparison, the quiescent galaxy and PSB mass profiles become relatively similar within $a = 3 \, kpc$ before reaching the region dominated by the PSF. One potential mechanism that can explain the evolution from the PSB profiles toward the quiescent galaxy profiles in Figure \ref{fig:stacked_profile} is ex-situ growth through dry minor mergers \citep[e.g.,][]{Bezanson.etal.2009,vanDokkum.etal.2010,Oser.etal.2010,Newman.etal.2012,Cheng.etal.2024}, in which the radius of the merged system can grow as the square of the change in mass when the mass of the infalling dwarf is insignificant in contrast to the main galaxy. This hypothesis is also favored by the fact that PSBs with apparent merger signatures appear to have larger light-weighted sizes than those without, as shown in the upper right panel of Figure \ref{fig:4panels} and Figure \ref{fig:mass_size}. The minor merger-driven ex-situ growth could have already started for a fraction of the PSBs in the first $1 \rm Gyr$ since quenching.

There are several caveats to the interpretation presented here. Firstly, we derive mass density profiles through the light distribution in only one filter for both PSBs (F110W) and quiescent galaxies (F125W) \footnote{Given the median redshift of this sample ($z_{median} = 1.15$), the difference in rest-frame effective wavelengths between F110W and F125W is small (less than $\rm 500\AA$) and the intrinsic color gradients between these two filters are negligible for our comparison.}, assuming there is no color gradient or change in mass-to-light ratio in the radial direction. For PSBs in this sample, this assumption is supported by several studies that find flat color gradients in some massive PSBs at intermediate redshifts \citep{Maltby.etal.2018,Suess.etal.2020,Setton.etal.2020}. In addition, such an assumption is self-consistent with the findings that there is no change in their relative sizes as they fade. The lack of growth suggests the stars that dominate the light from those galaxies are evenly distributed along the radial direction, consequently resulting in relatively flat color gradients. Meanwhile, both star-forming and quiescent galaxies tend to have color gradients that are bluer at large radii \citep{Mosleh.etal.2017, Maltby.etal.2018,Suess.etal.2019a,Suess.etal.2019b,Mosleh.etal.2020,Suess.etal.2021,theotherMiller.etal.2023}. These quiescent galaxies could have smaller mass-to-light ratios in the outskirts, and therefore overestimated mass density profile at larger radii. But the existence of negative color gradients in quiescent galaxies can be also explained by minor mergers, which would bring in younger ex-situ formed stellar populations in the outskirts on top of the flat color gradient of PSBs. The difference in color gradients between PSBs and typical star-forming galaxies implies that the progenitors of these PSBs are not among the typical star-forming galaxies if the central starburst was not the dominant evolutionary channel.

Secondly, the mass cut we implement in this analysis excludes the most massive PSBs and the majority of the subset with tidal features. The increased fraction of tidally disrupted features in those PSBs can scale up their mass density profile in the outskirts and compensate for the difference we see in Figure \ref{fig:stacked_profile}. However, we may have missed a potential tidally disrupted population in the low mass regime due to the detection limit in these images, if the surface brightness of disruption features scales with the overall brightness of these galaxies. 

Furthermore, the evolution pathways of galaxies have been shown to have complex branching structures \citep{Dubois.etal.2024}, reflecting the stochasticity in physical processes that drive their evolutionary behaviors. These PSBs may only represent a biased subset of the progenitors of quiescent galaxies. It is still possible that the descendants of these PSBs only make up the fraction of quiescent galaxies that are relatively compact. If the galaxies through other quenching channels join the quiescent population with larger sizes, they can increase the median size of the overall quiescent galaxy population. Alternatively, if some of these PSBs stay isolated from any merger events or rejuvenation, they can make up the population of red nugget relics discovered at lower redshift ($z<1$; \citealp{Lisiecki.etal.2023,Spiniello.etal.2024}) and in the local Universe \citep{Grebol-Tomas.etal.2023}, given their similar locations on the mass-size plane. To fully assess the complicated evolutionary pathways of fast-quenching galaxies and disentangle them from progenitor biases, a robust study of the number densities of these galaxies as a function of cosmic time is required. As the DESI SV LRG sample selection is incomplete at $z>0.8$, we are currently unable to robustly measure the number density of these PSBs \citep{Setton.etal.2023} and draw connections between them and any lower redshift descendant candidates. 

\section{Conclusion}

To shed light on the fast-quenching pathway of galaxy evolution, we spectroscopically select and study a sample of 171 massive ($\rm log(M_{*}/M_{\odot}) \sim 11)$ PSBs at $1<z<1.3$ from DESI LRG. Using HST WFC3/F110W imaging, we fit these galaxies with single S\'ersic profiles with the Bayesian fitting framework \texttt{pysersic}. We obtain robust estimates for their sizes as well as other parametric structural measures, such as S\'ersic index and axis ratio. This represents an order of magnitude increase in the number of spectroscopically confirmed PSBs with robust structural measurements above $z>1$. Combining these results with stellar population properties inferred from SED fitting via \texttt{Prospector} and comparing them with those of quiescent and star-forming galaxies in 3D-HST survey, we reach the following conclusions:
\begin{itemize}
\item{We robustly demonstrate that massive post-starburst galaxies are extremely compact at $1<z<1.3$. At similar stellar mass, their half-light radii are systematically $\rm \sim 0.1 \, dex$ smaller than those of quiescent galaxies and $\rm \sim 0.4 \, dex$ smaller than those of star-forming galaxies. }

\item{PSBs in this sample have central surface mass densities similar to quiescent galaxies ($\rm log(\Sigma_{1kpc}) \sim 10.1$), suggesting that the inner structures of PSBs are not significantly affected by any potential evolutionary mechanism after quenching. }

\item{These PSBs are round in projection ($b/a_{median} \sim 0.8$), suggesting that they are primarily spheroids, not disks, in 3D. }

\item{We find no trends in the median PSB sizes and central densities relative to their quiescent counterparts ($\rm \Delta log(R_{e})$ and $\rm \Delta log(\Sigma_{1kpc})$), within the time since both the major and minor bursts of star formation. This suggests that the previous episode of star formation was not centrally concentrated, but occurred throughout the galaxy.}
\end{itemize}

The PSB number density increases with redshift and the majority of the PSB population is located at earlier times \citep{Wild.etal.2016,Rowlands.etal.2018,Clausen.etal.2024}, suggesting that the rapid quenching path played a more dominant role in galaxy evolution in the early universe. Hence, it is necessary to find and characterize PSBs at the highest redshift possible. It is difficult to acquire large spectroscopic samples at higher redshifts, where objects are more significantly dimmed and require longer spectral integration time. The sample selection of DESI Validation Sample LRG, which enabled this work, is limited by a z-band magnitude cut of $\rm z_{fiber}<21.6$ and biased towards brighter objects \citep{Zhou.etal.2023} at $\rm z>0.8$. In the future, the Prime Focus Spectrograph \citep{Takada.etal.2014} surveys would enable us to spectroscopically select PSB samples that are magnitude complete and to estimate PSB number densities robustly at $z>1$.  In addition to this work, various samples at $\rm 0.5<z<2$ \citep{Yano.etal.2016,Almaini.etal.2017,Belli.etal.2019,Setton.etal.2022,Clausen.etal.2024} have reported PSB median sizes that are consistently smaller than median quiescent galaxy sizes at fixed mass in \cite{vanderWel.etal.2014}. But it is unclear whether this trend will continue at higher redshifts ($\rm z>2$). To study the sizes and structures of these rare objects at high redshifts, we need space-based images with a wavelength coverage redder than HST and a survey area wider than JWST. The Roman space telescope, which is designed to survey large areas at long wavelengths, would be ideal to provide IR images with high resolution and stable PSF for studying the morphology of high-z PSBs. Combining these data in future studies, we will be able to characterize PSBs around the cosmic noon and constrain the fast quenching path of galaxy evolution at earlier cosmic times.

\section*{Acknowledgements}
RB acknowledges support from the Research Corporation for Scientific Advancement (RCSA) Cottrell Scholar Award ID No: 27587 and from the National Science Foundation NSF-AAG grant \#1907697 and NSF-CAREER grant \# 2144314.

D.S. gratefully acknowledges the support provided by The Brinson Foundation through a Brinson Prize Fellowship grant for this work. The published results were also funded by the Polish National Agency for Academic Exchange (Bekker grant BPN/BEK/ 2021/1/00298/DEC/1) and the European Union's Horizon 2020 research and innovation program under the Maria SkłodowskaCurie (grant agreement No. 754510).

The data points used to create the figures in this work are available at 10.5281/zenodo.12693678.

This material is based upon work supported by the U.S. Department of Energy (DOE), Office of Science, Office of High-Energy Physics, under Contract No. DE–AC02–05CH11231, and by the National Energy Research Scientific Computing Center, a DOE Office of Science User Facility under the same contract. Additional support for DESI was provided by the U.S. National Science Foundation (NSF), Division of Astronomical Sciences under Contract No. AST-0950945 to the NSF’s National Optical-Infrared Astronomy Research Laboratory; the Science and Technology Facilities Council of the United Kingdom; the Gordon and Betty Moore Foundation; the Heising-Simons Foundation; the French Alternative Energies and Atomic Energy Commission (CEA); the National Council of Humanities, Science and Technology of Mexico (CONAHCYT); the Ministry of Science, Innovation and Universities of Spain (MICIU/AEI/10.13039/501100011033), and by the DESI Member Institutions: \url{https://www.desi.lbl.gov/collaborating-institutions}. Any opinions, findings, and conclusions or recommendations expressed in this material are those of the author(s) and do not necessarily reflect the views of the U. S. National Science Foundation, the U. S. Department of Energy, or any of the listed funding agencies.

The authors are honored to be permitted to conduct scientific research on Iolkam Du’ag (Kitt Peak), a mountain with particular significance to the Tohono O’odham Nation.

This research is based on observations made with the NASA/ESA Hubble Space Telescope obtained from the Space Telescope Science Institute, which is operated by the Association of Universities for Research in Astronomy, Inc., under NASA contract  NAS 5–26555. These observations are associated with the program HST/SNAP 17110.

The specific observations analyzed can be accessed via \dataset[https://doi.org/10.17909/scn6-hs39]{https://doi.org/10.17909/scn6-hs39}.

Support for program HST/SNAP 17110 was provided by NASA through a grant from the Space Telescope Science Institute, which is operated by the Association of Universities for Research in Astronomy, Inc., under NASA contract NAS 5–26555.

\facilities{
Mayall (DESI), 
HST (WFC3)
}

\software{Astropy \citep{astropy:2013,astropy:2018,astropy:2022}, AstroDrizzle \citep{AstroDrizzle}, GALFIT \citep{GALFIT}, Scipy \citep{2020SciPy}, SEP \citep{Bertin.etal.1996,Barbary2016}, Photutils \citep{photutils}, Prospector\citep{Leja.etal.2017,Johnson.etal.2021}, pysersic \citep{pysersic}}
 
\bibliography{sample631}{}
\bibliographystyle{aasjournal}

\appendix 

\section{Cutouts of selected galaxies in this sample}\label{appendix: all_gallery}
\begin{figure*}[h!tb]
    \centering
    \includegraphics[width = 0.97\textwidth]{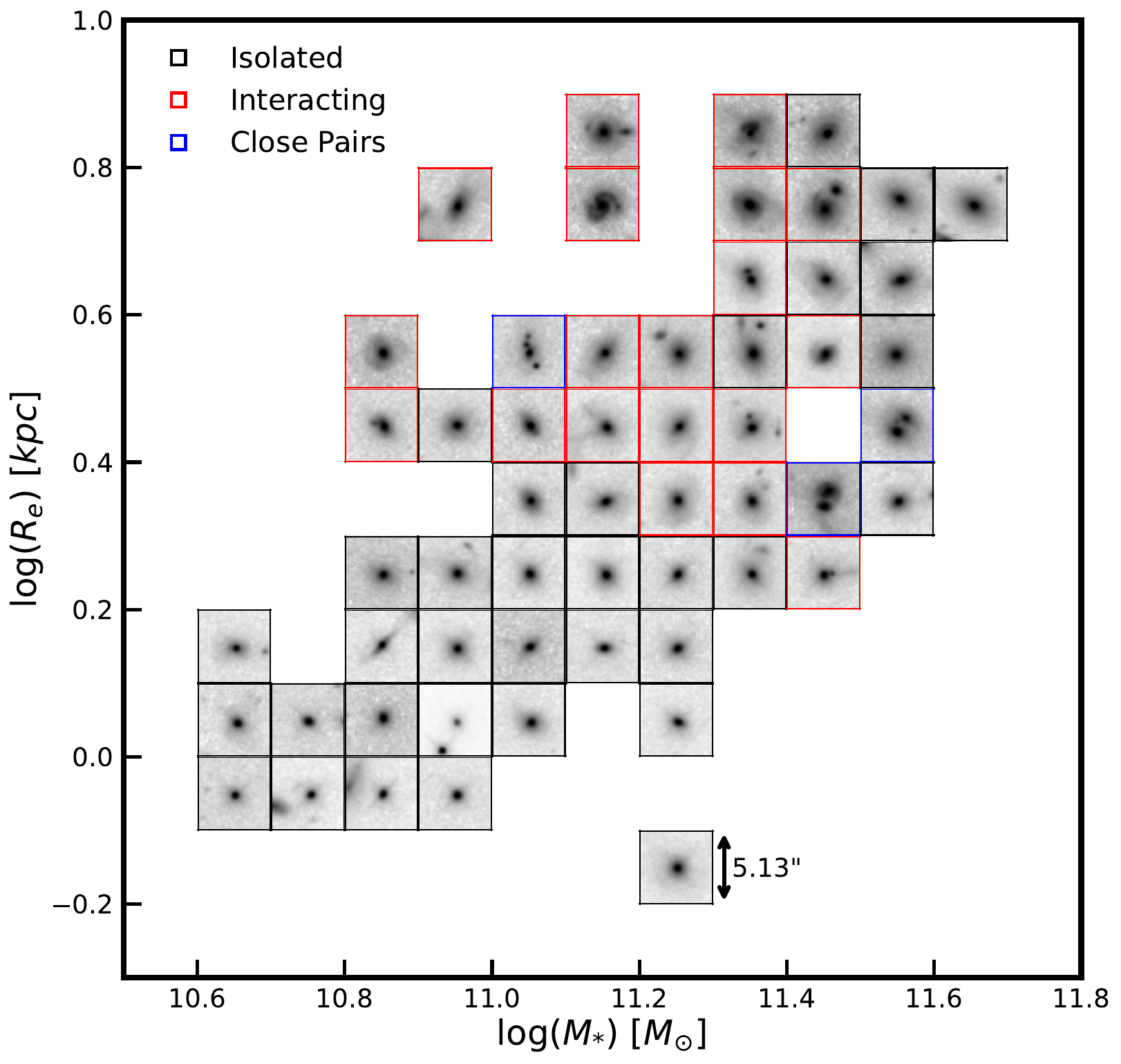}
    \caption{Selected HST WFC3/F110W image cutouts of PSB galaxies in this sample. Each field of view is centered on the object.}
    \label{fig:all_cutouts}
\end{figure*}

In Figure \ref{fig:all_cutouts}, we present a gallery of cutouts for selected galaxies in this sample. Each cutout is randomly selected from the ensemble of objects occupying the same parameter space in the size-mass plane. Cutouts are arranged according to their location in the PSB mass-size relation. We re-scale the pixel values in each image through an algorithm that mimics the implementation of log color scale in DS9, to better visualize the structures with low surface brightness. Galaxies identified as ``interacting" are framed in red, ``close pairs" are framed in blue, and ``isolated" are in black.

\section{Systematic differences between structural parameters fitted with \texttt{Galfit} and \texttt{pysersic}}\label{appendix: method_comparison}
\begin{figure*}[h!tb]
    \centering
    \includegraphics[width = 0.70\textwidth]{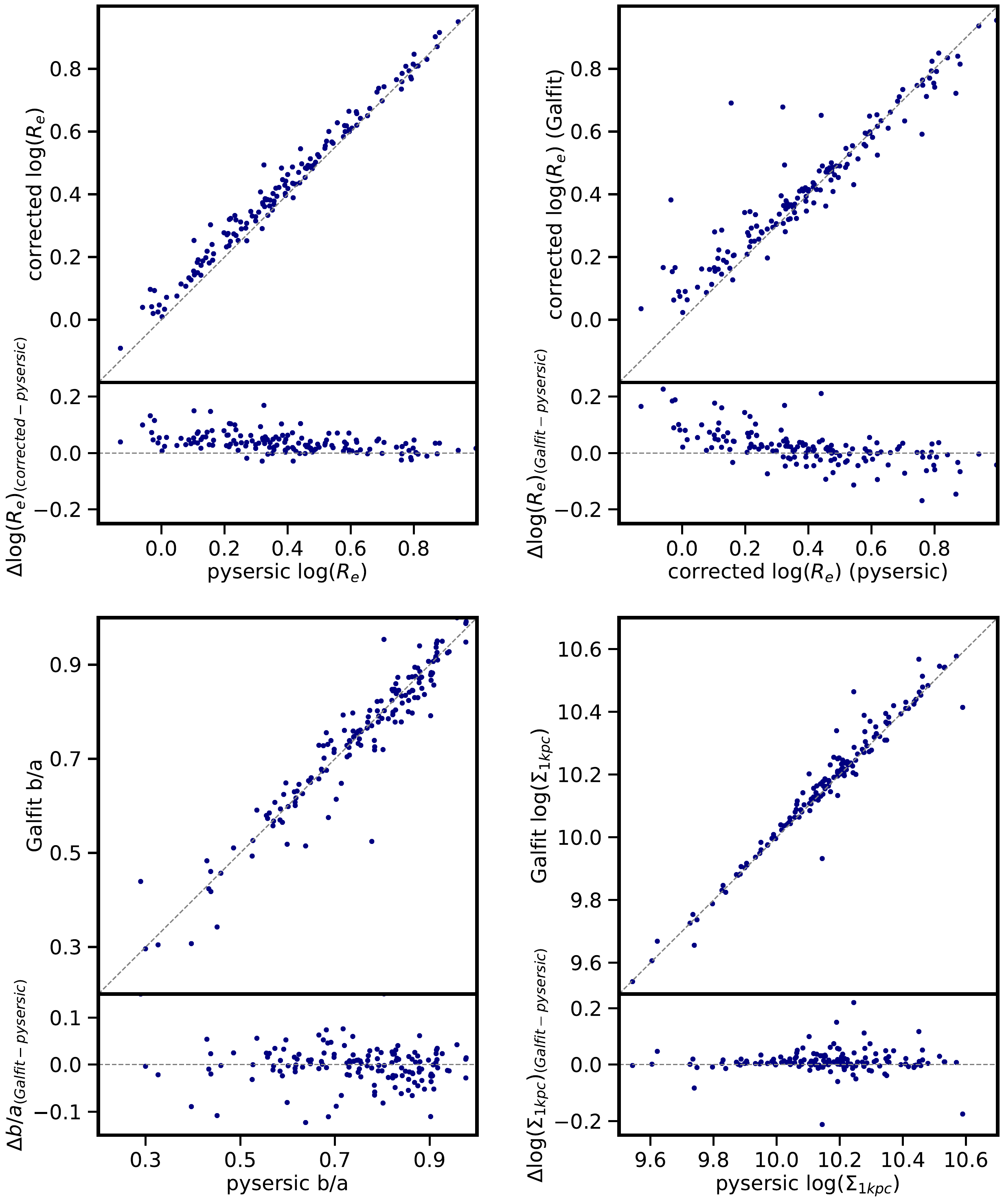}
    \caption{ (Upper left) Comparison between between the $R_{e}$ measured directly through \texttt{pysersic} and the $R_{e}$ corrected with residuals. In addition, we show the systematic differences in measured $R_{e}$ (Upper right), axis ratios (Lower left), and derived $\Sigma_{1kpc}$ (Lower right) that are obtained from \texttt{Galfit} and \texttt{pysersic}.}
    \label{fig:Appendix_B}
\end{figure*}

We show the systematic impact on half-light radii when applying the residual correction in \citep{Szomoru.etal.2012} in the upper left panel of Figure \ref{fig:Appendix_B}, using \texttt{pysersic} results. Correcting the half-light radii with residuals overall increases them by 0.03 dex in $\rm log(R_{e}) $. We compare the systematic differences between the residual-corrected half-light radii (upper right), axis ratios (lower left), and the derived central surface mass densities (lower right) inferred via \texttt{Galfit} and \texttt{pysersic}. The median value of half-light radii (residual-corrected) via \texttt{Galfit} is 0.01 dex in $\rm log(R_{e}) $ higher than that of \texttt{pysersic}. The median value of axis ratios via \texttt{Galfit} is almost identical to that of \texttt{pysersic}. The median value of central densities derived with parameters via \texttt{Galfit} is 0.01 dex higher than that of \texttt{pysersic}. Overall, the systematic difference between the results produced by \texttt{Galfit} and \texttt{pysersic} is small and has no significant impact on the qualitative conclusions in this work. The half-light radii measurements also do not deviate significantly after applying the residual correction and the deviations from S\'ersic profiles in these galaxies do not affect our conclusions qualitatively.

\end{document}